\magnification=\magstephalf
\newbox\SlashedBox 
\def\slashed#1{\setbox\SlashedBox=\hbox{#1}
\hbox to 0pt{\hbox to 1\wd\SlashedBox{\hfil/\hfil}\hss}{#1}}
\def\hboxtosizeof#1#2{\setbox\SlashedBox=\hbox{#1}
\hbox to 1\wd\SlashedBox{#2}}

% The following is necessary so that we can get a partial slash
% inside a math display... sigh.
\def\mathslashed#1{\setbox\SlashedBox=\hbox{$#1$}
\hbox to 0pt{\hbox to 1\wd\SlashedBox{\hfil/\hfil}\hss}#1}

\def\ifsmall{\iffalse}  % default is unreduced.
\def\titlepagefont{}  % default is ordinary font.

% the ps: landscape must be the first special command in order
% to get the first page in landscape mode -- so we go through some
% contortions to define TeXgraphics in the default case.
\def\DefineTeXgraphics{%
\special{ps::[global] /TeXgraphics { } def}}  % No need to do anything

\def\today{\ifcase\month\or January\or February\or March\or April\or May
\or June\or July\or August\or September\or October\or November\or
December\fi\space\number\day, \number\year}
\def\eatPrefix19{}
\def\Year{\expandafter\eatPrefix\the\year}
\newcount\hours \newcount\minutes
\def\monthname{\ifcase\month\or
January\or February\or March\or April\or May\or June\or July\or
August\or September\or October\or November\or December\fi}
\def\shortmonthname{\ifcase\month\or
Jan\or Feb\or Mar\or Apr\or May\or Jun\or Jul\or
Aug\or Sep\or Oct\or Nov\or Dec\fi}

\def\TimeStamp{\hours\the\time\divide\hours by60%
\minutes -\the\time\divide\minutes by60\multiply\minutes by60%
\advance\minutes by\the\time%
${\rm \shortmonthname}\cdot\if\day<10{}0\fi\the\day\cdot\the\year%
\qquad\the\hours:\if\minutes<10{}0\fi\the\minutes$}

%\DefineTeXgraphics}

%\DefineTeXgraphics}

%\DefineTeXgraphics}

\def\Title#1{%
\vskip 1in{\titlefont\centerline{#1}}\vskip .5in}
%\DefineTeXgraphics}
 
\def\Date#1{\leftline{#1}\tenrm\supereject%
\global\hsize=\hsbody\global\hoffset=\hbodyoffset%
\footline={\hss\tenrm\folio\hss}}% restores pagenumbers

\newif\ifdraftmode
\newif\ifleftlabels  % Labels in left margins as well, for European-size paper

% Stolen from harvmac.tex 04/08/92
%       use \nolabels to get rid of eqn, ref, and fig labels in draft mode
\def\nolabels{\def\wrlabeL##1{}\def\eqlabeL##1{}\def\reflabeL##1{}}
\def\writelabels{\def\wrlabeL##1{\leavevmode\vadjust{\rlap{\smash%
{\line{{\escapechar=` \hfill\rlap{\sevenrm\hskip.03in\string##1}}}}}}}%
\def\eqlabeL##1{{\escapechar-1\rlap{\sevenrm\hskip.05in\string##1}}}%
\def\reflabeL##1{\noexpand\rlap{\noexpand\sevenrm[\string##1]}}}
\def\writeleftlabels{\def\wrlabeL##1{\leavevmode\vadjust{\rlap{\smash%
{\line{{\escapechar=` \hfill\rlap{\sevenrm\hskip.03in\string##1}}}}}}}%
\def\eqlabeL##1{{\escapechar-1%
\rlap{\sixrm\hskip.05in\string##1}%
\llap{\sevenrm\string##1\hskip.03in\hbox to \hsize{}}}}%
\def\reflabeL##1{\noexpand\rlap{\noexpand\sevenrm[\string##1]}}}
\nolabels

\input hyperbasics.tex

\newdimen\fullhsize
\newdimen\hstitle
\hstitle=\hsize % default
\newdimen\hsbody
\hsbody=\hsize % default
\newdimen\hbodyoffset
\hbodyoffset=\hoffset % default
\newbox\leftpage
\def\abstract#1{#1}
\def\rotated{\special{ps: landscape}
\magnification=1000  % This line must come before we change vsize,
                     % since \magnification sets it to a fixed value.
\baselineskip=14pt
\global\hstitle=9truein\global\hsbody=4.75truein
\global\vsize=7truein\global\voffset=-.31truein
\global\hoffset=-0.54in\global\hbodyoffset=-.54truein
\global\fullhsize=10truein
\def\DefineTeXgraphics{%
\special{ps::[global] 
/TeXgraphics {currentpoint translate 0.7 0.7 scale
              -80 0.72 mul -1000 0.72 mul translate} def}}
 % 0.7 is slightly less than the ratio of horizontal sizes: 4.75 to 6.5
\let\lr=L
\def\ifsmall{\iftrue}
\def\titlepagefont{\twelvepoint}
\trueseventeenpoint
\def\almostshipout##1{\if L\lr \count1=1
      \global\setbox\leftpage=##1 \global\let\lr=R
   \else \count1=2
      \shipout\vbox{\hbox to\fullhsize{\box\leftpage\hfil##1}}
      \global\let\lr=L\fi}

\output={\ifnum\count0=1 %%% This is the HUTP version
 \shipout\vbox{\hbox to \fullhsize{\hfill\pagebody\hfill}}\advancepageno
 \else
 \almostshipout{\leftline{\vbox{\pagebody\makefootline}}}\advancepageno 
 \fi}

\def\abstract##1{{\leftskip=1.5in\rightskip=1.5in ##1\par}} }

% Messages on lines by themselves
\def\linemessage#1{\immediate\write16{#1}}

% tagged sec numbers
\global\newcount\secno \global\secno=0
\global\newcount\appno \global\appno=0
\global\newcount\meqno \global\meqno=1
\global\newcount\subsecno \global\subsecno=0
% and figure numbers
\global\newcount\figno \global\figno=0

\newif\ifAnyCounterChanged
% If we are comparing numbers, there's no special problem.
% But if we are comparing roman numerals, we must be careful, because
% stuff read in from the aux file would be made up of ordinary
% characters (category code = 11), whereas \romannumeral generates
% characters with category code = 12..., so the stuff from the
% current run won't appear equal to the previous definition, as far
% as \warnIfChanged is concerned.
% To get around this, we have a macro \makeNormal, which converts
% letters `ivxlcdmIVXLCDM' to normal letters, no matter what their category
% code.  The macro has the convoluted form it does, with aftergroup's & all,
% to avoid blowing up TeX...
% The macro is used below in makeNormalizedRomappno, by which means we
% define the appendix counters to be strings containing vanilla versions
% of the letters... Sigh
\let\terminator=\relax
% The string to be normalized must not contain { and } tokens...
\def\normalize#1{\ifx#1\terminator\let\next=\relax\else%
\if#1i\aftergroup i\else\if#1v\aftergroup v\else\if#1x\aftergroup x%
\else\if#1l\aftergroup l\else\if#1c\aftergroup c\else%
\if#1m\aftergroup m\else%
\if#1I\aftergroup I\else\if#1V\aftergroup V\else\if#1X\aftergroup X%
\else\if#1L\aftergroup L\else\if#1C\aftergroup C\else%
\if#1M\aftergroup M\else\aftergroup#1\fi\fi\fi\fi\fi\fi\fi\fi\fi\fi\fi\fi%
\let\next=\normalize\fi%
\next}
% makes #1 a normalized version of #2...
\def\makeNormal#1#2{\def\doNormalDef{\edef#1}\begingroup%
\aftergroup\doNormalDef\aftergroup{\normalize#2\terminator\aftergroup}%
\endgroup}
% makes a normalized version of its argument:

\def\warnIfChanged#1#2{%
\ifundef#1% skip it
\else\begingroup%
\edef\oldDefinitionOfCounter{#1}\edef\newDefinitionOfCounter{#2}%
%\message{old: \oldDefinitionOfCounter}%
%\message{new: \newDefinitionOfCounter}%
\ifx\oldDefinitionOfCounter\newDefinitionOfCounter%
\else%
\linemessage{Warning: definition of \noexpand#1 has changed.}%
\global\AnyCounterChangedtrue\fi\endgroup\fi}

\def\Section#1{\global\advance\secno by1\relax\global\meqno=1%
\global\subsecno=0%
\bigbreak\bigskip% (combination \goodbreak\bigskip\bigskip)
\centerline{\twelvepoint \bf %
\the\secno. #1}%
\par\nobreak\medskip\nobreak}
\def\tagsection#1{%
\warnIfChanged#1{\the\secno}%
\xdef#1{\the\secno}%
\ifWritingAuxFile\immediate\write\auxfile{\noexpand\xdef\noexpand#1{#1}}\fi%
}
\def\section{\Section}
\def\Subsection#1{\global\advance\subsecno by1\relax\medskip %
\leftline{\bf\the\secno.\the\subsecno\ #1}%
\par\nobreak\smallskip\nobreak}
\def\tagsubsection#1{%
\warnIfChanged#1{\the\secno.\the\subsecno}%
\xdef#1{\the\secno.\the\subsecno}%
\ifWritingAuxFile\immediate\write\auxfile{\noexpand\xdef\noexpand#1{#1}}\fi%
}

\def\subsection{\Subsection}

\def\romappno{\uppercase\expandafter{\romannumeral\appno}}
\def\makeNormalizedRomappno{%
\expandafter\makeNormal\expandafter\normalizedromappno%
\expandafter{\romannumeral\appno}%
\edef\normalizedromappno{\uppercase{\normalizedromappno}}}
\def\Appendix#1{\global\advance\appno by1\relax\global\meqno=1\global\secno=0%
\global\subsecno=0%
\bigbreak\bigskip% (combination \goodbreak\bigskip\bigskip)
\centerline{\twelvepoint \bf Appendix %
\romappno. #1}%
\par\nobreak\medskip\nobreak}
\def\tagappendix#1{\makeNormalizedRomappno%
\warnIfChanged#1{\normalizedromappno}%
\xdef#1{\normalizedromappno}%
\ifWritingAuxFile\immediate\write\auxfile{\noexpand\xdef\noexpand#1{#1}}\fi%
}
\def\appendix{\Appendix}
\def\Subappendix#1{\global\advance\subsecno by1\relax\medskip %
\leftline{\bf\romappno.\the\subsecno\ #1}%
\par\nobreak\smallskip\nobreak}
\def\tagsubappendix#1{\makeNormalizedRomappno%
\warnIfChanged#1{\normalizedromappno.\the\subsecno}%
\xdef#1{\normalizedromappno.\the\subsecno}%
\ifWritingAuxFile\immediate\write\auxfile{\noexpand\xdef\noexpand#1{#1}}\fi%
}

\def\eqn#1{\makeNormalizedRomappno%
\ifnum\secno>0%
  \warnIfChanged#1{\the\secno.\the\meqno}%
  \eqno(\the\secno.\the\meqno)\xdef#1{\the\secno.\the\meqno}%
     \global\advance\meqno by1
\else\ifnum\appno>0%
  \warnIfChanged#1{\normalizedromappno.\the\meqno}%
  \eqno({\rm\romappno}.\the\meqno)%
      \xdef#1{\normalizedromappno.\the\meqno}%
     \global\advance\meqno by1
\else%
  \warnIfChanged#1{\the\meqno}%
  \eqno(\the\meqno)\xdef#1{\the\meqno}%
     \global\advance\meqno by1
\fi\fi%
\eqlabeL#1%
\ifWritingAuxFile\immediate\write\auxfile{\noexpand\xdef\noexpand#1{#1}}\fi%
}
\def\defeqn#1{\makeNormalizedRomappno%
\ifnum\secno>0%
  \warnIfChanged#1{\the\secno.\the\meqno}%
  \xdef#1{\the\secno.\the\meqno}%
     \global\advance\meqno by1
\else\ifnum\appno>0%
  \warnIfChanged#1{\normalizedromappno.\the\meqno}%
  \xdef#1{\normalizedromappno.\the\meqno}%
     \global\advance\meqno by1
\else%
  \warnIfChanged#1{\the\meqno}%
  \xdef#1{\the\meqno}%
     \global\advance\meqno by1
\fi\fi%
\eqlabeL#1%
\ifWritingAuxFile\immediate\write\auxfile{\noexpand\xdef\noexpand#1{#1}}\fi%
}
\def\anoneqn{\makeNormalizedRomappno%
\ifnum\secno>0
  \eqno(\the\secno.\the\meqno)%
     \global\advance\meqno by1
\else\ifnum\appno>0
  \eqno({\rm\normalizedromappno}.\the\meqno)%
     \global\advance\meqno by1
\else
  \eqno(\the\meqno)%
     \global\advance\meqno by1
\fi\fi%
}
\def\mfig#1#2{\ifx#20%unnumbered figure
\else\global\advance\figno by1%
\relax#1\the\figno%
\warnIfChanged#2{\the\figno}%
\xdef#2{\the\figno}%
\reflabeL#2%
\ifWritingAuxFile\immediate\write\auxfile{\noexpand\xdef\noexpand#2{#2}}\fi\fi%
}

\def\fig#1{\mfig{fig.\ ~}#1}

\catcode`@=11 % borrow the private macros of PLAIN (with care)

% \LoadFigure is used to put a figure into the text.  Its first argument
% is the symbolic name for the figure (if it isn't defined, a new number
% will be assigned);  the second argument is a caption;
% the third argument size information in the form 
% \epsfxsize=3.0in\epsfysize=3.5in (this argument may be blank and
% may contain any valid preparatory argument used by the epsf package);
% the fourth and last argument is the name of the file which contains the 
% figure.
% The macro is basically just a front-end for \epsfbox; its purpose is 
% to allow figures to be switched from placement in the running text
% to placement on a separate page at the end of the text.  This choice
% is made using the flag \FiguresInText{true,false}; in the latter case,
% figures are placed at the end, size information is ignored (figures
% will be full-size), and the captions are listed separately on a page
% when the \listfigs command is invoked, followed by the figures, each
% on a separate page.  
%  The epsf package must be loaded by the user.
%  To change the size of captions in the text, redefine \captionsize.
\newif\ifFiguresInText\FiguresInTexttrue
\newif\if@FigureFileCreated
\newwrite\capfile
\newwrite\figfile

%default
\newif\ifcaption
\captiontrue
\def\captionsize{\tenrm}
\def\PlaceTextFigure#1#2#3#4{%
\vskip 0.5truein%
#3\hfil\epsfbox{#4}\hfil\break%
\ifcaption\hfil\vbox{\captionsize Figure #1. #2}\hfil\fi%
\vskip10pt}
\def\PlaceEndFigure#1#2{%
\epsfxsize=\hsize\epsfbox{#2}\vfill\centerline{Figure #1.}\eject}

\def\LoadFigure#1#2#3#4{%
\ifundef#1{\phantom{\mfig{}#1}}\else%  Write out definition only if it's new.
% Bop figure counter...
%\ifx#10% unnumbered figure
%\else\warnIfChanged#1{\the\figno}%
%\ifWritingAuxFile\immediate\write\auxfile{\noexpand\xdef\noexpand#1{#1}}\fi%\fi
\fi%
\ifFiguresInText% Figure is immediate
\PlaceTextFigure{#1}{#2}{#3}{#4}%
\else% Figure is at the end
\if@FigureFileCreated\else%
\immediate\openout\capfile=\jobname.caps%
\immediate\openout\figfile=\jobname.figs%
@FigureFileCreatedtrue\fi%
\immediate\write\capfile{\noexpand\item{Figure \noexpand#1.\ }{#2}\vskip10pt}%
\immediate\write\figfile{\noexpand\PlaceEndFigure\noexpand#1{\noexpand#4}}%
\fi}

\def\listfigs{\ifFiguresInText\else%
\vfill\eject\immediate\closeout\capfile%\parindent=20pt
\immediate\closeout\figfile%
\centerline{{\bf Figures}}\bigskip\frenchspacing%
\catcode`@=11 % borrow the private macros of PLAIN (with care)
\def\captionsize{\tenrm}
\input \jobname.caps\vfill\eject\nonfrenchspacing%
\catcode`\@=\active
\catcode`@=12  % No longer.
\input\jobname.figs\fi}

%\font\titlefont=cmr10 at 16pt
\font\ninerm=cmr9
\font\eightrm=cmr8
\font\sixrm=cmr6

\def\loadtrueseventeenpoint{
 \font\seventeenrm=cmr10 at 17.28truept
 \font\seventeeni=cmmi10 at 17.28truept
 \font\seventeenbf=cmbx10 at 17.28truept
 \font\seventeenit=cmti10 at 17.28truept
 \font\seventeensl=cmsl10 at 17.28truept
 \font\seventeensy=cmsy10 at 17.28truept
}
\def\loadfourteenpoint{
\font\fourteenrm=cmr10 at 14.4pt
\font\fourteeni=cmmi10 at 14.4pt
\font\fourteenit=cmti10 at 14.4pt
\font\fourteensl=cmsl10 at 14.4pt
\font\fourteensy=cmsy10 at 14.4pt
\font\fourteenbf=cmbx10 at 14.4pt
}
\def\loadtruetwelvepoint{
\font\twelverm=cmr10 at 12truept
\font\twelvei=cmmi10 at 12truept
\font\twelveit=cmti10 at 12truept
\font\twelvesl=cmsl10 at 12truept
\font\twelvesy=cmsy10 at 12truept
\font\twelvebf=cmbx10 at 12truept
}

\font\ninei=cmmi9
\font\eighti=cmmi8
\font\sixi=cmmi6
\skewchar\ninei='177 \skewchar\eighti='177 \skewchar\sixi='177

\font\ninesy=cmsy9
\font\eightsy=cmsy8
\font\sixsy=cmsy6
\skewchar\ninesy='60 \skewchar\eightsy='60 \skewchar\sixsy='60

\font\ninebf=cmbx9
\font\eightbf=cmbx8
\font\sixbf=cmbx6

\font\ninett=cmtt9
\font\eighttt=cmtt8

\hyphenchar\tentt=-1 % inhibit hyphenation in typewriter type
\hyphenchar\ninett=-1
\hyphenchar\eighttt=-1         

\font\ninesl=cmsl9
\font\eightsl=cmsl8

\font\nineit=cmti9
\font\eightit=cmti8
\font\sevenit=cmti7

% Added 07/09/97, initially, 10-point...
\scriptfont\itfam=\sevenit

 % unslanted text italic
                      
\newskip\ttglue
\def\tenpoint{\def\rm{\fam0\tenrm}%
  \textfont0=\tenrm \scriptfont0=\sevenrm \scriptscriptfont0=\fiverm
  \textfont1=\teni \scriptfont1=\seveni \scriptscriptfont1=\fivei
  \textfont2=\tensy \scriptfont2=\sevensy \scriptscriptfont2=\fivesy
  \textfont3=\tenex \scriptfont3=\tenex \scriptscriptfont3=\tenex
  \def\it{\fam\itfam\tenit}%
      \textfont\itfam=\tenit\scriptfont\itfam=\sevenit
  \def\sl{\fam\slfam\tensl}\textfont\slfam=\tensl
  \def\bf{\fam\bffam\tenbf}\textfont\bffam=\tenbf \scriptfont\bffam=\sevenbf
  \scriptscriptfont\bffam=\fivebf
  \normalbaselineskip=12pt
  \let\sc=\eightrm
  \let\big=\tenbig
  \setbox\strutbox=\hbox{\vrule height8.5pt depth3.5pt width\z@}%
  \normalbaselines\rm}

\def\twelvepoint{\def\rm{\fam0\twelverm}%
  \textfont0=\twelverm \scriptfont0=\ninerm \scriptscriptfont0=\sevenrm
  \textfont1=\twelvei \scriptfont1=\ninei \scriptscriptfont1=\seveni
  \textfont2=\twelvesy \scriptfont2=\ninesy \scriptscriptfont2=\sevensy
  \textfont3=\tenex \scriptfont3=\tenex \scriptscriptfont3=\tenex
  \def\it{\fam\itfam\twelveit}\textfont\itfam=\twelveit
  \def\sl{\fam\slfam\twelvesl}\textfont\slfam=\twelvesl
  \def\bf{\fam\bffam\twelvebf}\textfont\bffam=\twelvebf%
  \scriptfont\bffam=\ninebf
  \scriptscriptfont\bffam=\sevenbf
  \normalbaselineskip=12pt
  \let\sc=\eightrm
  \let\big=\tenbig
  \setbox\strutbox=\hbox{\vrule height8.5pt depth3.5pt width\z@}%
  \normalbaselines\rm}

\def\fourteenpoint{\def\rm{\fam0\fourteenrm}%
  \textfont0=\fourteenrm \scriptfont0=\tenrm \scriptscriptfont0=\sevenrm
  \textfont1=\fourteeni \scriptfont1=\teni \scriptscriptfont1=\seveni
  \textfont2=\fourteensy \scriptfont2=\tensy \scriptscriptfont2=\sevensy
  \textfont3=\tenex \scriptfont3=\tenex \scriptscriptfont3=\tenex
  \def\it{\fam\itfam\fourteenit}\textfont\itfam=\fourteenit
  \def\sl{\fam\slfam\fourteensl}\textfont\slfam=\fourteensl
  \def\bf{\fam\bffam\fourteenbf}\textfont\bffam=\fourteenbf%
  \scriptfont\bffam=\tenbf
  \scriptscriptfont\bffam=\sevenbf
  \normalbaselineskip=17pt
  \let\sc=\elevenrm
  \let\big=\tenbig                                          
  \setbox\strutbox=\hbox{\vrule height8.5pt depth3.5pt width\z@}%
  \normalbaselines\rm}

\def\seventeenpoint{\def\rm{\fam0\seventeenrm}%
  \textfont0=\seventeenrm \scriptfont0=\fourteenrm \scriptscriptfont0=\tenrm
  \textfont1=\seventeeni \scriptfont1=\fourteeni \scriptscriptfont1=\teni
  \textfont2=\seventeensy \scriptfont2=\fourteensy \scriptscriptfont2=\tensy
  \textfont3=\tenex \scriptfont3=\tenex \scriptscriptfont3=\tenex
  \def\it{\fam\itfam\seventeenit}\textfont\itfam=\seventeenit
  \def\sl{\fam\slfam\seventeensl}\textfont\slfam=\seventeensl
  \def\bf{\fam\bffam\seventeenbf}\textfont\bffam=\seventeenbf%
  \scriptfont\bffam=\fourteenbf
  \scriptscriptfont\bffam=\twelvebf
  \normalbaselineskip=21pt
  \let\sc=\fourteenrm
  \let\big=\tenbig                                          
  \setbox\strutbox=\hbox{\vrule height 12pt depth 6pt width\z@}%
  \normalbaselines\rm}

\def\ninepoint{\def\rm{\fam0\ninerm}%
  \textfont0=\ninerm \scriptfont0=\sixrm \scriptscriptfont0=\fiverm
  \textfont1=\ninei \scriptfont1=\sixi \scriptscriptfont1=\fivei
  \textfont2=\ninesy \scriptfont2=\sixsy \scriptscriptfont2=\fivesy
  \textfont3=\tenex \scriptfont3=\tenex \scriptscriptfont3=\tenex
  \def\it{\fam\itfam\nineit}\textfont\itfam=\nineit
  \def\sl{\fam\slfam\ninesl}\textfont\slfam=\ninesl
  \def\bf{\fam\bffam\ninebf}\textfont\bffam=\ninebf \scriptfont\bffam=\sixbf
  \scriptscriptfont\bffam=\fivebf
  \normalbaselineskip=11pt
  \let\sc=\sevenrm
  \let\big=\ninebig
  \setbox\strutbox=\hbox{\vrule height8pt depth3pt width\z@}%
  \normalbaselines\rm}

\def\eightpoint{\def\rm{\fam0\eightrm}%
  \textfont0=\eightrm \scriptfont0=\sixrm \scriptscriptfont0=\fiverm%
  \textfont1=\eighti \scriptfont1=\sixi \scriptscriptfont1=\fivei%
  \textfont2=\eightsy \scriptfont2=\sixsy \scriptscriptfont2=\fivesy%
  \textfont3=\tenex \scriptfont3=\tenex \scriptscriptfont3=\tenex%
  \def\it{\fam\itfam\eightit}\textfont\itfam=\eightit%
  \def\sl{\fam\slfam\eightsl}\textfont\slfam=\eightsl%
  \def\bf{\fam\bffam\eightbf}\textfont\bffam=\eightbf \scriptfont\bffam=\sixbf%
  \scriptscriptfont\bffam=\fivebf%
  \normalbaselineskip=9pt%
  \let\sc=\sixrm%
  \let\big=\eightbig%
  \setbox\strutbox=\hbox{\vrule height7pt depth2pt width\z@}%
  \normalbaselines\rm}

 % use after $ in ninepoint sections
\def\tenbig#1{{\hbox{$\left#1\vbox to8.5pt{}\right.\n@space$}}}
\def\ninebig#1{{\hbox{$\textfont0=\tenrm\textfont2=\tensy
  \left#1\vbox to7.25pt{}\right.\n@space$}}}
\def\eightbig#1{{\hbox{$\textfont0=\ninerm\textfont2=\ninesy
  \left#1\vbox to6.5pt{}\right.\n@space$}}}

% Page layout
%\newinsert\footins
\def\footnote#1{\edef\@sf{\spacefactor\the\spacefactor}#1\@sf
      \insert\footins\bgroup\eightpoint
      \interlinepenalty100 \let\par=\endgraf
        \leftskip=\z@skip \rightskip=\z@skip
        \splittopskip=10pt plus 1pt minus 1pt \floatingpenalty=20000
        \smallskip\item{#1}\bgroup\strut\aftergroup\@foot\let\next}
\skip\footins=12pt plus 2pt minus 4pt % space added when footnote is present
%\count\footins=1000 % footnote magnification factor (1 to 1)
\dimen\footins=30pc % maximum footnotes per page

\newinsert\margin
\dimen\margin=\maxdimen
%\count\margin=0 \skip\margin=0pt % marginal inserts take up no space
\def\titlefont{\seventeenpoint}
\loadtruetwelvepoint % At FNAL...
\loadtrueseventeenpoint

% \use\cs
% puts in the expansion of `\cs' if it's defined, the literal "\cs" otherwise.
\def\eatOne#1{}
\def\ifundef#1{\expandafter\ifx%
\csname\expandafter\eatOne\string#1\endcsname\relax}
\def\notTrue{\iffalse}\def\isTrue{\iftrue}
\def\ifdef#1{{\ifundef#1%
\aftergroup\notTrue\else\aftergroup\isTrue\fi}}
\def\use#1{\ifundef#1\linemessage{Warning: \string#1 is undefined.}%
{\tt \string#1}\else#1\fi}

%     \ref\label{text}
% generates a number, assigns it to \label, generates an entry.
% To list the refs on a separate page,  \listrefs
% \nref does the same without generating any text at the reference
% point
% June 26 1994: \preref postpones the generation of an entry, along with
% the text, until the first use of the reference

% 09/14/95: Added html...
%\def\hyperref#1#2{\special{html:<a href=\quote#1\quote>}{#2}\special{html:</a>}}
% 09/25/95: Now using Tanmoy Battacharya's macros...

%
\catcode`"=11
\let\quote="
\catcode`"=12
\chardef\foo="22
\global\newcount\refno \global\refno=1
\newwrite\rfile
\newlinechar=`\^^J
\def\@ref#1#2{\the\refno\n@ref#1{#2}}
% Added 09/14/95{\the\refno\n@ref#1{#2}}
\def\h@ref#1#2#3{\href{#3}{\the\refno}\n@ref#1{#2}}
\def\n@ref#1#2{\xdef#1{\the\refno}%
\ifnum\refno=1\immediate\openout\rfile=\jobname.refs\fi%
\immediate\write\rfile{\noexpand\item{[\noexpand#1]\ }#2.}%
\global\advance\refno by1}
\def\nref{\n@ref} % Hide to allow redefinitions of \ref,\nref to \preref
\def\ref{\@ref}   % without breaking the latter...
\def\hrref{\h@ref}
% To start a long reference...
\def\lref#1#2{\the\refno\xdef#1{\the\refno}%
\ifnum\refno=1\immediate\openout\rfile=\jobname.refs\fi%
\immediate\write\rfile{\noexpand\item{[\noexpand#1]\ }#2\semi}%
\global\advance\refno by1}
% To continue a long reference...
\def\cref#1{\immediate\write\rfile{#1\semi}}
% To end a long reference...

\def\preref#1#2{\gdef#1{\@ref#1{#2}}}

\def\semi{;\hfil\noexpand\break}

\def\listrefs{\vfill\eject\immediate\closeout\rfile%\parindent=20pt
\centerline{{\bf References}}\bigskip\frenchspacing%
\input \jobname.refs\vfill\eject\nonfrenchspacing}

\def\inputAuxIfPresent#1{\immediate\openin1=#1
\ifeof1\message{No file \auxfileName; I'll create one.
}\else\closein1\relax\input\auxfileName\fi%
}
% For references, some journal names
\def\NPB{Nucl.\ Phys.\ B}

%and archives...

%\def\hepphref#1{\hyperref{http://xxx.lanl.gov/abs/hep-ph/#1}{archive}%
%{hep-ph/#1}{hep-ph/#1}}

%\def\hepphref#1{\href{http://xxx.lanl.gov/abs/hep-ph/#1}{hep-ph/#1}}

% An .aux file --- for forward references...
\newif\ifWritingAuxFile
\newwrite\auxfile
\def\SetUpAuxFile{%
\xdef\auxfileName{\jobname.aux}%
% Read it in if it exists
\inputAuxIfPresent{\auxfileName}%
% Now write a new one.
\WritingAuxFiletrue%
\immediate\openout\auxfile=\auxfileName}

% Some generally useful notation
\def\L{\left(}\def\R{\right)}
\def\LP{\left.}\def\RP{\right.}
\def\LB{\left[}\def\RB{\right]}

\def\RV{\right|}

% Warn about changed counters...
\def\bye{\par\vfill\supereject%
\ifAnyCounterChanged\linemessage{
Some counters have changed.  Re-run tex to fix them up.}\fi%
\end}

\catcode`\@=\active
\catcode`@=12  % No longer.
\catcode`\"=\active

\def\Tr{\mathop{\rm Tr}\nolimits}
\def\pol{\varepsilon}

\def\spa#1.#2{\left\langle#1\,#2\right\rangle}
\def\spb#1.#2{\left[#1\,#2\right]}
\SetUpAuxFile
\hfuzz 20pt
\overfullrule 0pt

\def\e{\epsilon}
\def\tree{{\rm tree\vphantom{p}}}
\def\oneloop{{\rm 1\hbox{\sevenrm-}loop}}
\def\twoloop{{\rm 2{\sevenrm-}loop}}
\def\lloop{\hbox{\sevenrm-}{\rm loop}}
\def\Split{\mathop{\rm Split}\nolimits}

\def\Ctree{\Split^\tree}
\def\Cone{\Split^\oneloop}
\def\Ctwo{\Split^\twoloop}
\def\tcdot{\mskip -1mu\cdot\mskip-1mu}
\def\ll#1{{\lambda_{#1}}}
\def\la{\ll{a}}
\def\lb{\ll{b}}
\def\lc{\ll{c}}

\def\llongrightarrow{%
\relbar\mskip-0.5mu\joinrel\mskip-0.5mu\relbar\mskip-0.5mu\joinrel\longrightarrow}
\def\inlimit^#1{\buildrel#1\over\llongrightarrow}

\def\LIPS{{\rm LIPS}}
\def\phpol{{\rm ph.\ pol.}}

\loadfourteenpoint

%%%%%%%%%%%%%%%%%%%%%%%%%%%%%%%%%%%%%%%%%%%%%%%%%%%%%%%%%
\input epsf

\noindent\nopagenumbers
hep-ph/9901201 \hfill{Saclay/SPhT--T98/144}

\leftlabelstrue
\vskip -1.0 in
\Title{All-Order Collinear Behavior in Gauge Theories}

\baselineskip17truept
\centerline{David A. Kosower}
\baselineskip12truept
\centerline{\it Service de Physique Th\'eorique${}^{\natural}$}
\centerline{\it Centre d'Etudes de Saclay}
\centerline{\it F-91191 Gif-sur-Yvette cedex, France}
\centerline{\tt kosower@spht.saclay.cea.fr}

\vskip 0.2in\baselineskip13truept

\vskip 0.5truein
\centerline{\bf Abstract}

{\narrower 
I give a proof, using the unitarity-based
method, of the collinear factorization of
the leading-color contribution 
to gauge-theory amplitudes.   
The proof also provides a concrete formula which can be used
to compute the associated splitting amplitudes.

}
\vskip 0.3truein

\centerline{\it Submitted to Nuclear Physics B}

\vfill
\vskip 0.1in
\noindent\hrule width 3.6in\hfil\break
\noindent
${}^{\natural}$Laboratory of the
{\it Direction des Sciences de la Mati\`ere\/}
of the {\it Commissariat \`a l'Energie Atomique\/} of France.\hfil\break

\Date{}

\line{}

\baselineskip17pt
%

%%%%%%%%%%%%%%%%%%%%%%%%%%%%%%%%%%%%%%%%%%%%%%%%%%%%%%%%%%%%%%%%
%  Refs.
\preref\Color{%
F. A. Berends and W. T. Giele,
Nucl.\ Phys.\ B294:700 (1987)\semi
D. A.\ Kosower, B.-H.\ Lee and V. P. Nair, Phys.\ Lett.\ 201B:85 (1988)\semi
M.\ Mangano, S. Parke and Z.\ Xu, Nucl.\ Phys.\ B298:653 (1988)\semi
Z. Bern and D. A.\ Kosower, Nucl.\ Phys.\ B362:389 (1991)}

\preref\GG{W.T.\ Giele and E.W.N.\ Glover,
Phys.\ Rev.\ D46:1980 (1992)}
\preref\GGK{W.T.\ Giele, E. W. N.\ Glover and D. A. Kosower,
Nucl.\ Phys.\ B403:633 (1993) [hep-ph/9302225]}

\preref\Recurrence{F. A. Berends and W. T. Giele, Nucl.\ Phys.\ B306:759 (1988)\semi
D. A. Kosower, Nucl.\ Phys.\ B335:23 (1990)}

\preref\ManganoReview{%
M. Mangano and S.J. Parke, Phys.\ Rep.\ 200:301 (1991)}

\preref\CS{%
S. Catani and M. Seymour, Phys.\ Lett.\ B378:287 (1996) [hep-ph/9602277];
Nucl.\ Phys.\ B485:291 (1997) [hep-ph/9605323]}

\preref\CST{S. Catani, M. H. Seymour, and Z. Tr\'ocs\'anyi,
       Phys.\ Rev.\ D55:6819 (1997) [hep-ph/9610553]}

\preref\SoftGluonReview{A. Bassetto, M. Ciafaloni, and G. Marchesini, 
 Phys.\ Rep.\ 100:201 (1983)}
\preref\AP{G. Altarelli and G. Parisi, Nucl.\ Phys.\ B126:298 (1977)}
\preref\FKS{S. Frixione, Z. Kunszt, and A. Signer, 
   Nucl.\ Phys.\ B467:399 (1996) [hep-ph/9512328]}

\preref\Byckling{E. Byckling and K. Kajantie, {\it Particle Kinematics\/}
(Wiley, 1973)}

\preref\KobaNielsenR{Z. Koba and H. B. Nielsen, Nucl.\ Phys.\ B10:633 (1969)\semi
J. H. Schwarz, Phys.\ Rep.\ 89:223 (1982)}

\preref\CDR{J. C.\ Collins, {\it Renormalization} (Cambridge University Press, 1984)}
\preref\HV{G. 't Hooft and M. Veltman, Nucl.\ Phys.\ B44:189 (1972)}
\preref\DimRed{W. Siegel, Phys.\ Lett.\ 84B:193 (1979)\semi
D.M.\ Capper, D.R.T.\ Jones and P. van Nieuwenhuizen, Nucl.\ Phys.\
B167:479 (1980)\semi
L.V.\ Avdeev and A.A.\ Vladimirov, Nucl.\ Phys.\ B219:262 (1983)}
\preref\FDHS{Z. Bern and D. A.\ Kosower, \NPB 379:451 (1992)}

\preref\BernChalmers{Z. Bern and G. Chalmers, Nucl.\ Phys.\ B447:465 (1995) 
  [hep-ph/9503236]}
\preref\Pentagon{Z. Bern, L. Dixon, and D. A. Kosower,
            Nucl.\ Phys.\ B412:751 (1994) [hep-ph/9306240]}

\preref\SusyOne{Z. Bern, L. Dixon, D. C. Dunbar and D. A. Kosower, 
        Nucl.\ Phys.\ B435:59 (1995) [hep-ph/9409265]}
\preref\SusyFour{Z. Bern, L. Dixon, D. C. Dunbar, and D. A. Kosower,
Nucl.\ Phys.\ B425:217 (1994) [hep-ph/9403226]}
\preref\qqggg{Z. Bern, L. Dixon, and D. A. Kosower,
Nucl.\ Phys.\  B437:259 (1995) [hep-ph/9409393]}
\preref\AnnRev{Z. Bern, L. Dixon, and D. A. Kosower,
Ann.\ Rev.\ Nucl.\ Part.\ Sci.\ 46 109 (1996) [hep-ph/9602280]}

%% Double collinear
\preref\EnglishDoubleCollinear{J. M. Campbell and E. W. N. Glover,
Nucl.\ Phys.\ B527:264 (1998) [hep-ph/9710255]}
\preref\ItalianDoubleCollinear{S. Catani and M. Grazzini, preprint hep-ph/9810389}

\section{Introduction}
\vskip 10pt

Gauge-theory amplitudes have simple and universal 
features in various singular limits.
In the collinear limit, when the momenta
of two massless particles become proportional
($k_a\rightarrow z (k_a+k_b)$,
$k_b\rightarrow (1-z) (k_a+k_b)$), 
we may summarize these factorization properties
in a set of universal {\it splitting amplitudes\/}.
The squares of the tree-level collinear
splitting amplitudes, summed over helicities,
yield the
Altarelli--Parisi kernels~[\use\AP,\use\SoftGluonReview].  Their integrals
 play an important role in the infrared
cancellations that are needed for calculations of
physical quantities at next-to-leading order in
perturbative quantum chromodynamics~[\use\GG,\use\FKS,\use\CS,\use\GGK].

Beyond leading order, splitting amplitudes also
have relatively simple forms.  Their explicit
forms at one loop~[\use\SusyFour,\use\qqggg] were extracted by taking the
collinear limit in various five-point one-loop amplitudes.
They are useful as complements to the unitarity-based
method of computing one-loop amplitudes~[\use\SusyOne,\use\AnnRev], and
will play a role in the cancellation of infrared
divergences in next-to-next-to-leading order calculations.

Bern and Chalmers~[\use\BernChalmers] have previously given a proof
of collinear factorization at one loop.  It has the
virtue of being quite general, and not limited to
gauge theories.  It is unclear, however, how to
generalize their proof beyond one loop.  Furthermore,
even at one loop
it provides neither a simple understanding of
the structure of the splitting amplitudes, nor
a simple method of computing them.  The purpose
of this paper is to provide an independent proof
of collinear factorization which generalizes readily
to higher loops.  It will also provide a simple and concrete
formula for computing the splitting amplitudes at one
and two loops.  The proof is an application of the
unitarity-based method for computing loop amplitudes
developed by Bern, Dixon, Dunbar, and the author~[\use\SusyOne,\use\AnnRev].

\section{Order of Limits}
\tagsection\LimitOrderSection
\vskip 10pt

The properties of non-Abelian gauge-theory amplitudes
in the collinear limit are easiest to express
in the context of a color decomposition~[\use\Color].  For 
tree-level all-gluon amplitudes in an $SU(N)$ gauge theory 
the color decomposition has the form,
$$
{\cal A}_n^\tree(\{k_i,\lambda_i,a_i\}) = 
\sum_{\sigma \in S_n/Z_n} \Tr(T^{a_{\sigma(1)}}\cdots T^{a_{\sigma(n)}})\,
A_n^\tree(\sigma(1^{\lambda_1},\ldots,n^{\lambda_n}))\,,
\eqn\ColorDecomposition$$
where $S_n/Z_n$ is the group of non-cyclic permutations
on $n$ symbols, and $j^{\lambda_j}$ denotes the $j$-th momentum
and helicity.  As is by now standard,
I use the normalization $\Tr(T^a T^b) = \delta^{ab}$,
and I suppress powers of the coupling constant.
One can write analogous formul\ae\ for amplitudes
with quark-antiquark pairs or uncolored external lines.
The color-ordered or partial amplitude $A_n$ is gauge invariant.

\LoadFigure\BasicFigure
{\baselineskip 13 pt
\noindent\narrower\ninerm
The discontinuity of a loop amplitude expressed as a product of tree amplitudes.
}  {\epsfxsize 2.2 truein}{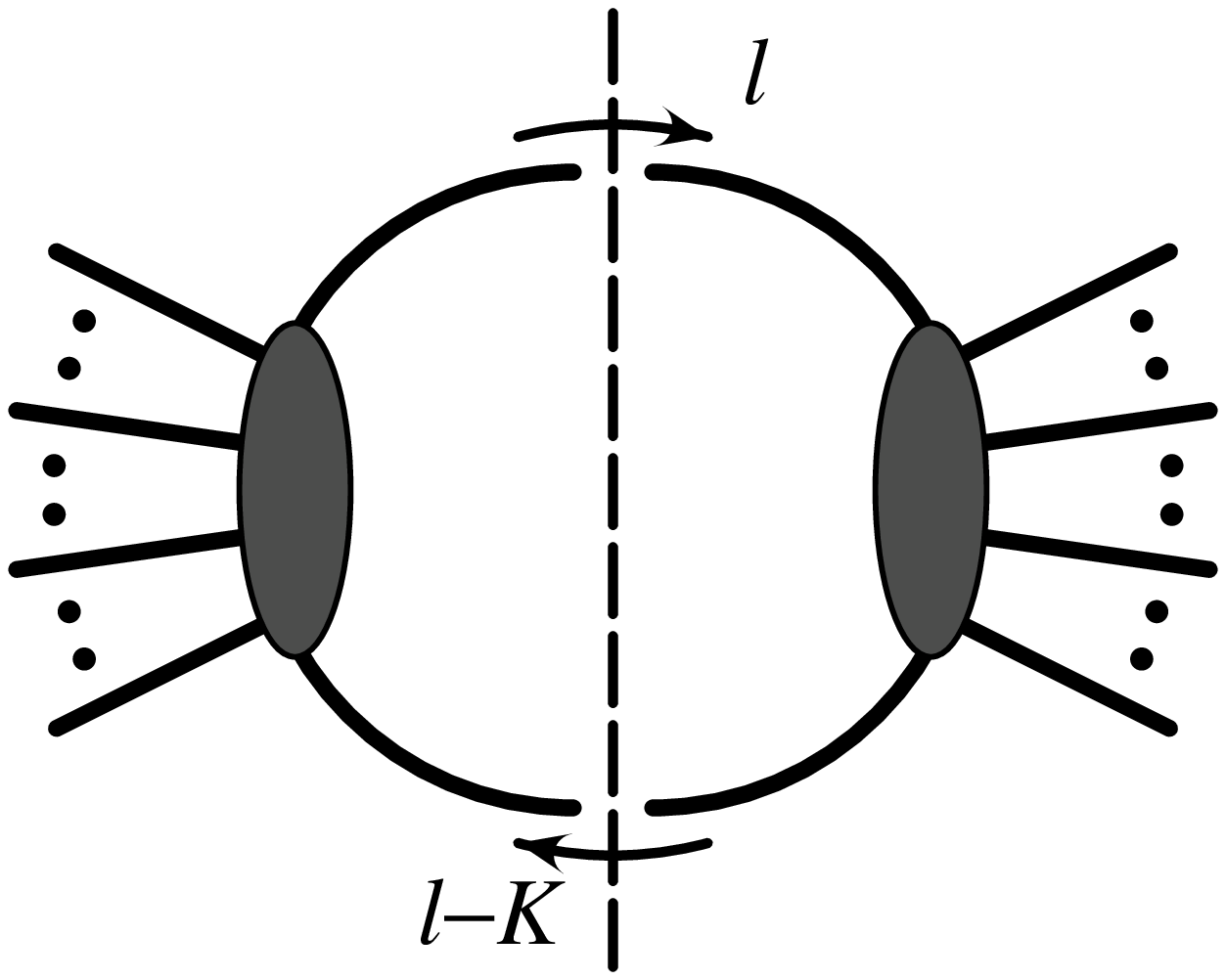}{}

In the cut-based approach to one-loop calculations, we have the following
basic equation,
$$\eqalign{
A^\oneloop &= \sum_{{\rm cuts\ } K^2} 
\int {d^{4-2\e} \ell\over (2\pi)^{4-2\e}} 
   {i\over \ell^2} A^\tree_{\rm left} {i\over (\ell-K)^2} A^\tree_{\rm right}
}\eqn\BasicCut$$
where the loop integral is promoted from the integral over phase space,
and the sum over all cuts is understood to count only once integral
functions that have cuts in multiple channels.  The right-hand side is depicted in 
fig.~\use\BasicFigure.
(I will leave the
sum over all intermediate particle types and physical polarizations
implicit in the cut expression.
In the four-dimensional helicity scheme~[\use\FDHS,\use\CST], these would
be the positive and negative helicity states; in the conventional
dimensional regularization scheme~[\use\CDR], it would include `$\e$'
helicities as well.)  

In general, we are interested only in the resulting terms through
finite order in the dimensional regularization parameter
$\e$.  Reconstructing the amplitude by summing over the cuts does not however
commute with expanding around $\e=0$; only if all cuts satsify a 
power-counting criterion can the full amplitude, to order $\e^0$, be
reconstructed from the $D=4$ tree amplitudes on either side of the cut.
Otherwise, we must keep the cuts to higher orders in $\e$ (or use other
tricks) in order to obtain the finite rational terms in the amplitude.

The singular limit in which we are interested poses additional 
interchange-of-limits problems.  The limit in which $k_a \parallel k_b$,
or more generally in which a single invariant $s_{ab} = (k_a+k_b)^2$ of
two color-adjacent legs vanishes while all other invariants stay
(roughly) fixed, fails to commute with the expansion around $\e=0$,
as can be seen in the example,
$$
\lim_{s\rightarrow 0} {s^{-\e}\over\e^2} = 0 \neq \lim_{s\rightarrow0}
 {1\over\e^2} - {\ln s\over\e} + {\ln^2 s\over 2}\,.
\anoneqn$$
(As is conventional in investigating infrared divergences, I will
take $\e<0$.)
This interchange-of-limits issue in specific integral functions
is addressed by the Bern--Chalmers discontinuity functions [\BernChalmers].

\def\onemass{{\rm 1m}}
\def\rg{r_\Gamma}
\def\Li{\mathop{\rm Li}\nolimits}
\def\Ord{{\cal O}}
\def\F#1#2{\,{{\vphantom{F}}_{#1}F_{#2}}}
In what order should we take the singular and four-dimensional limits?
It might seem more natural to take the $\e\rightarrow 0$ limit first,
but this is not the best choice.
The problem is precisely that in this order, we lose
control not only of divergent terms but also of
finite terms in the integrals (the $\ln^2 s$ above).
 Keeping them under control would mean keeping
track of all finite terms in integrals, a rather formidable task.
On the other hand, if we take the singular limit first, then all
one-loop integrals with more than one external invariant (i.e. all
but one-mass triangles and bubbles) are in fact {\it finite\/} in
any singular limit.  This may seem bizarre --- what about the $s t$
in the denominator of the one-mass box, for example? --- but is easily
seen to be true using the Feynman parameter representation.  (The resulting
collinear-limit integral may, of course, be more singular in the later
$\e\rightarrow 0$ limit, but that is a later concern.)  Indeed, it
is instructive to consider the $s,t\rightarrow0$ limit of the 
reduced one-mass
box~[\use\Pentagon]; what naively looks to be uncontrolled growth
the standard $\e\rightarrow 0$
expansion,
$$\eqalign{
I_4^\onemass(s,t,m^2)\ &=\ {\rg \over st} \biggl\{
  {2\over\e^2} \Bigl[ 
1 + \e\ln \Bigl({\mu^2\over -s}\Bigr)
+ \e\ln \Bigl({\mu^2\over -t}\Bigr)
- \e\ln \Bigl({\mu^2\over -m^2}\Bigr)\Bigl]\cr
&\ 
+ \ln^2\Bigl({\mu^2\over -s}\Bigr)
+ \ln^2\Bigl({\mu^2\over -t}\Bigr)
+ \ln^2\Bigl({\mu^2\over -m^2}\Bigr)\cr
 &\ -\ 2\ \Li_2\left(1-{m^2\over s}\right)
  \ -\ 2\ \Li_2\left(1-{m^2\over t}\right)
  \ -\ \ln^2\left({s\over t}\right)\ -\ {\pi^2\over3} \biggr\}
  \ +\ \Ord(\e), \cr
}\eqn\NaiveLimit$$
(with $\rg \equiv\ {\Gamma(1+\e)\Gamma^2(1-\e)/\Gamma(1-2\e)}$)
turns out to be quite well behaved (after analytic continuation)
if we go back to the Feynman parameter
representation,
$$\eqalign{
 (\mu^2)^\e\Gamma(2+\e)
\int_0^1 d^4a\ 
{\delta(1 - {\textstyle \sum_i} a_i)\over
  \LB -s a_1 a_3 - t a_2 a_4 - m^2 a_4 a_1\RB^{2+\e}}
&\longrightarrow (\mu^2)^\e\Gamma(2+\e)
\int_0^1 d^4a\ 
{\delta(1 - {\textstyle \sum_i} a_i)\over \LB - m^2 a_4 a_1\RB^{2+\e}}\;.
\cr &= -2\rg  (\mu^2)^\e (-m^2)^{-2-\e} 
}\eqn\FPlimit$$
The analytic continuation involved, however, is not appropriate for us,
as it amounts to subtracting poles in $s$ and $t$.  (The scale $\mu$ is
introduced by dimensional regularization and would eventually become
the ${\overline {\rm MS}}$ renormalization scale.)

We do want to pick up such poles, so the appropriate limit for
our purposes is given by expanding the integrated form,
leaving epsilonic powers alone, and taking care
to drop only terms suppressed by
{\it integer\/} powers of $s/m^2$ or $t/m^2$.

Using the form of $I_4^\onemass$ expressed in terms of hypergeometric
functions,
$$\eqalign{
&{2\rg\over\e^2\, s t} \; \Biggl[
     \smash{\L {s t\over \mu^2(t-m^2)}\R^{-\e}}
   \F21\L-\e,-\e;1-\e; {s+t-m^2\over t-m^2}\R\cr
&\hskip 7mm
    + \L {s t\over s-m^2}\R^{-\e} \F21\L -\e,-\e;1-\e;
         {s+t-m^2\over \mu^2(s-m^2)}\R\cr
&\hskip 7mm -\smash{\L {s t m^2\over (s+t-m^2) m^2 - s t}\R^{-\e}}
  \F21\L -\e,-\e;1-\e;
      {(s + t-m^2) m^2\over \mu^2[(s+t-m^2) m^2 - s t]}\R
     \Biggr]\,,\cr
}\anoneqn$$
dropping terms suppressed by integer powers of (say) $s/m^2$, 
and only then expanding in $\e$, we find
$$\eqalign{
&{2\rg\over\e^2} \; 
     \L {s t\over -m^2\mu^2}\R^{-\e}
   \F21\L-\e,-\e;1-\e; 1\R = 
{2\rg\over\e^2} \; 
     \L {s t\over -m^2\mu^2}\R^{-\e} \Gamma(1-\e)\Gamma(1+\e)\cr
&= {2\rg\over\e^2} \; 
     \L {(-s) (-t)\over -m^2\mu^2}\R^{-\e} + {\pi^2\over 3} + \Ord(\e)\,,\cr
}\eqn\ExpandTwo$$
which is well-behaved in the singular limit.  We should thus take the
singular limit first, and only then the four-dimensional one.
In the following discussion, unless otherwise stated,
I will then keep amplitudes and cut expressions to all orders in $\e$,
unexpanded in a series.

Of course, there is no free lunch.  The discrepancy between
eqn.~(\use\FPlimit) and~(\use\ExpandTwo) hints at the problem that
naive manipulations with the Feynman-parameter representation dispose
of terms like $s^{-\e}$, which we really want to keep.  If we are
considering a cut in one channel, say $s_1$, and the resulting cut is
proportional to $s_2^{-\e}$, where $s_2\rightarrow 0$ in the desired
singular limit, then it would appear at first glance that
this contribution will
disappear quietly, and we won't pick it up.  (Of course, if the
$s_2^{-\e}$ does not show up as part of an integral function
containing a cut in the $s_1$ channel, but merely as a separate
function from the integral reductions, then it should be dropped
anyway because the cut in the $s_1$ channel necessarily has an
ambiguity of that form.)  However, this contribution will {\it also\/}
have a cut in the $s_2$ channel, and we can expect to pick it up
there.  We will therefore have to be very careful about taking
the limit of the cut in the singular channel.
  This subtlety would also mean 
in considering singular limits with more than one singular invariant,
that we 
would have to take cuts in multiple channels simultaneously, in order to
pick up contributions proportional to $s_1^{-\e}
s_2^{-\e}$.

While we cannot take the singular limit before computing the cut
for the cut in the singular channel itself, what about other channels?
In general, an inverse integer power of the singular invariant must
be supplied by the amplitudes on either side of the cut,
or else by the loop integral whose discontinuity is given by the cut.
(The helicity algebra eventually softens this singularity to a square-root
from a full pole.)  If the inverse integer power is supplied by
an amplitude on either side of the cut, we can clearly take the
singular limit before considering the cut.  The same turns out
to be true (except in the singular channel, as discussed above)
even if the inverse integer power arises from the loop integral
itself.  Because we are working to all orders in
$\e$, the integral will necessarily have a cut in the singular channel.
Thus whatever mistake we would be making by taking the limit
naively in some other channel will be fixed up when we consider
the cut in the singular channel itself.  This implies, in particular,
that cuts in which the collinear momenta are on opposite sides
of the cut can be dropped.  Such configurations {\it do\/} give
contributions in the collinear limit, for example in the form
of the hard two-mass box integral; but
this contribution can be picked up in the singular channel.

\section{Collinear Factorization at One Loop}
\vskip 10pt

At one loop, all amplitudes can be written as sums of leading-color
amplitudes\footnote*{Or primitive amplitudes, when some of the external
legs are fermions.}.  It is thus sufficient to consider the
collinear limit of leading-color amplitudes in order to characterize
the complete collinear behavior of a one-loop amplitude.
Furthermore, the amplitude is finite in the limit unless the
two collinear legs are (color-) adjacent, so restrict attention to
this case.

The amplitude can be reconstructed completely from the knowledge
of all cuts; because we are working to all orders in $\e$, there
is no possible rational ambiguity.  If we
can determine the collinear limits of all cuts, we will then have determined
the collinear limit of the amplitude.

Consider first a generic cut, in which the two collinear legs $a$
and $b$ are a proper subset of the legs on one side of the cut,
and label the legs on the other side from $c$ through $d$.
The cut in this channel, $t_{c\ldots d} = (k_c+\cdots+k_d)^2$, is then
$$\eqalign{
&\LP A_n^{\oneloop}(1,\ldots,a,b,\ldots,n) \RV_{t_{c\cdots d} {\rm\ cut}}
=\cr
&\hskip 10mm \int d\LIPS^{4-2\e}(\ell_1,-\ell_2)\;
   A_{n-m+2}^\tree(\ell_1,c,\ldots,d,-\ell_2) 
      A_{m+2}^\tree(\ell_2,d+1,\ldots,a,b,\ldots,c-1,-\ell_1)\cr
}\eqn\typeNS$$

To obtain the collinear limits of this expression, we must make
use of the tree-level factorization,
$$\eqalign{
&A_{n}^\tree(1,\ldots,a^\la,b^\lb,\ldots,n) 
   \inlimit^{k_a \tcdot k_b\rightarrow 0}
%\cr &\hskip 15mm
 \sum_{\phpol\ \sigma}
 \Ctree_{-\sigma}(a^\la,b^\lb) 
    A_{n-1}^\tree(1,\ldots,(a+b)^\sigma,\ldots,n)\,,\cr
}\eqn\CollinearA$$
where $\Ctree$ is the usual tree splitting amplitude, 
the notation `$a+b$' means $k_a+k_b$, 
and where the
notation `$\phpol$' indicates a sum over physical polarizations only.
$\Ctree(a,b)$ goes as $s_{ab}^{-1/2}$; terms not singular in the
limit are omitted in this factorization.  This factorization is
depicted schematically in \fig\TreeFactorizationFigure.

This form may be derived from either the Berends--Giele recurrence
relations~[\use\Recurrence], or else~[\use\ManganoReview] 
from the Koba--Nielsen open-string amplitude~[\use\KobaNielsenR].
Either derivation shows that this factorization holds only for 
on-shell (that is, physically polarized) legs $a,b$, but
in arbitrary dimension.  The following arguments will thus go
through equally well in the four-dimensional helicity scheme,
the conventional dimensional regularization scheme, or 
the original 't~Hooft--Veltman scheme.

\LoadFigure\TreeFactorizationFigure
{\baselineskip 13 pt
\noindent\narrower\ninerm
A schematic depiction of the collinear factorization of tree-level
amplitudes, with the amplitudes labelled clockwise.
}  {\epsfxsize 4.2 truein}{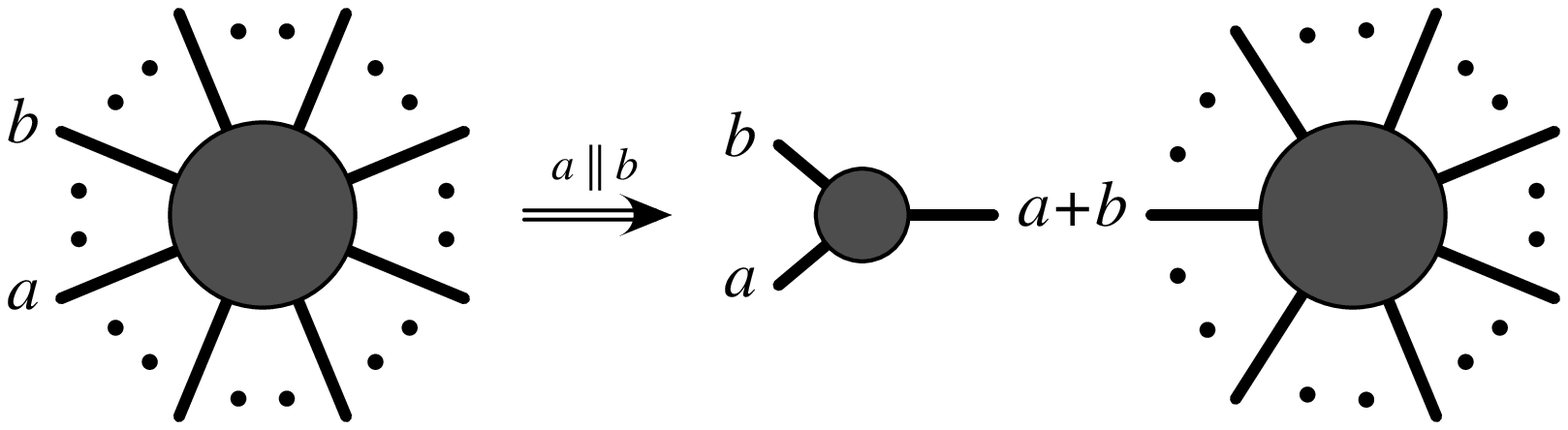}{}

The tree splitting amplitude is given by the
appropriate limit of the three-point Berends-Giele current,
$$\eqalign{
\Ctree_{\sigma}(a^\la,b^\lb) &= 
{1\over \sqrt{2} s_{ab}} \LB 
  \pol_a^{(\la)}\tcdot\pol_b^{(\lb)}\,(k_b-k_a)\tcdot\pol_{-\Sigma}^{(\sigma)}
  + 2 k_a\tcdot\pol_b^{(\lb)}\,\pol_a^{(\la)}\tcdot\pol_{-\Sigma}^{(\sigma)}
  - 2 k_b\tcdot\pol_a^{(\la)}\,\pol_b^{(\lb)}\tcdot\pol_{-\Sigma}^{(\sigma)}
\RB\,,\cr
}\anoneqn$$
where $\Sigma$ denotes the fused leg, $k_\Sigma = k_a+k_b$.

In the limit, eqn.~(\use\typeNS) then yields,
$$\eqalign{
&\sum_{\phpol\ \sigma} \Ctree_{-\sigma}(a^\la,b^\lb)\,
   \int d\LIPS^{4-2\e}(\ell_1,-\ell_2)\;
   \cr
&\hskip 15mm\times
   A_{n-m+2}^\tree(\ell_1,c,\ldots,d,-\ell_2) 
      A_{m+1}^\tree(\ell_2,d\!+\!1,\ldots,(a+b)^\sigma,\ldots,c\!-\!1,-\ell_1)\cr
&= \sum_{\phpol\ \sigma} \Ctree_{-\sigma}(a^\la,b^\lb)\,
\LP A_{n-1}^{\oneloop}(1,\ldots,(a+b)^\sigma,\ldots,n) \RV_{t_{c\cdots d} {\rm\ cut}} \,.
\cr}\eqn\CollinearB$$

As noted in section~\LimitOrderSection, we need not consider
cuts where the momenta are on opposite sides of the cut (in which
case they are both necessarily adjacent to it).  The above
derivation breaks down, as expected, if $a$ and $b$ are the
only legs on one side of the cut; but all contributions except those
detectable in the singular channel take the form presented in
eqn.~(\use\CollinearB).  This leaves us with the singular channel, which I
consider next.

\section{The Singular Channel}
\tagsection\OneLoopSingularSection
\vskip 10pt

The cut in the singular channel has the following form,
$$\eqalign{
&\LP A_n^{\oneloop}(1,\ldots,a,b,\ldots,n) \RV_{s_{ab} {\rm\ cut}}
=\cr
&\hskip 10mm \sum_{\phpol \lambda_{1,2}} \int d\LIPS^{4-2\e}(\ell_1,\ell_2)\;
   A_{n}^\tree(-\ell_1^{-\ll1},b\!+\!1,\ldots,a\!-\!1,-\ell_2^{-\ll2}) 
      A_{4}^\tree(\ell_2^\ll2,a,b,\ell_1^\ll1)\,.\cr
}\eqn\typeS$$
(I have relabelled $\ell_1\rightarrow-\ell_1$ in this equation.)
We already know we must treat it carefully, in particular we
are {\it not\/} allowed to take the singular limit before 
performing the cut integral.  It is nonetheless reassuring to
see that the expression itself raises several warning flags,
independent of the discussion in section~\use\LimitOrderSection.
The cut integral is somewhat peculiar, because the volume of phase
space vanishes in the singular limit.  Furthermore, the factorization
we used in the previous section fails. We cannot factorize
the four-point amplitude into splitting amplitudes times three-point
amplitudes, because the three-point amplitude vanishes on-shell.
(Note, incidentally, that this does {\it not\/} mean that the
four-point amplitude is not singular in this limit; on the
contrary, it is {\it more\/} singular, having a full pole rather
than just a square-root singularity.  This is of course just
the usual forward-scattering singularity of a gauge theory.)

The other special feature of this cut is what allows us to
simplify this expression.  In this cut, momentum conservation 
conservation forces $s_{\ell_1\ell_2}\longrightarrow 0$ in the
limit.  This in turn allows us to use the tree-level 
factorization~(\use\CollinearA) on the amplitude on the left-hand
side of the cut~(\use\typeS),
$$\eqalign{
   &A_{n}^\tree(-\ell_1^{-\ll1},b+1,\ldots,a-1,-\ell_2^{-\ll2}) 
\inlimit^{s_{ab}\rightarrow 0}\cr
&\hskip 15mm
 \sum_{\phpol\ \sigma} 
   \Ctree_{-\sigma}(-\ell_2^{-\ll2},-\ell_1^{-\ll1})\,
   A_{n-1}^\tree((-\ell_1-\ell_2)^\sigma,b+1,\ldots,a-1) \cr
&\hskip 15mm= \sum_{\phpol\ \sigma} 
   \Ctree_{-\sigma}(-\ell_2^{-\ll2},-\ell_1^{-\ll1})\,
   A_{n-1}^\tree((a+b)^\sigma,b+1,\ldots,a-1)\,. \cr
}\anoneqn$$
Note that the only remaining dependence on the cut momenta
is in the tree-level splitting function.
\LoadFigure\OneLoopSplittingFigure
{\baselineskip 13 pt
\noindent\narrower\ninerm
The defining equation for the one-loop splitting amplitude.
}  {\epsfxsize 4.2 truein}{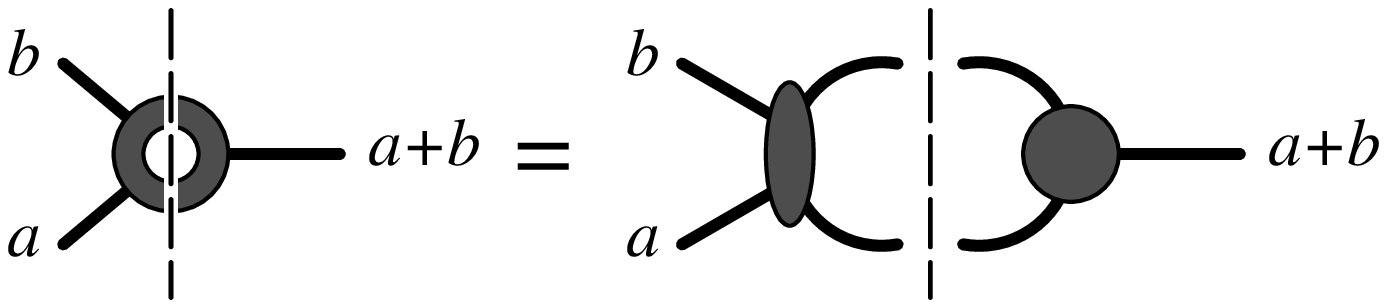}{}

The cut in eqn.~(\use\typeS) thus becomes
$$\eqalign{
&\sum_{{\phpol\atop\sigma,\ll1,\ll2}} 
     A_{n-1}^\tree((a+b)^\sigma,b\!+\!1,\ldots,a\!-\!1)\,
\cr &\hskip 20mm\times
\int d\LIPS^{4-2\e}(\ell_1,\ell_2)\;
   \Ctree_{-\sigma}(-\ell_2^{-\ll2},-\ell_1^{-\ll1})\,
      A_{4}^\tree(\ell_2^\ll2,a^\la,b^\lb,\ell_1^\ll1)\cr
&\equiv \sum_{\phpol\ \sigma} A_{n-1}^\tree((a+b)^\sigma,b,\ldots,a-1)\,
\LP\Cone_{-\sigma}(a^\la,b^\lb)\RV_{s_{ab} \rm\, cut}\,\cr
}\eqn\typeSb$$
in the collinear limit.  Since this channel is the only one in which
$\Cone$ has a cut, we can immediately write down a formula for it,
$$\eqalign{
\Cone_{-\sigma}&(a^\la,b^\lb) = \cr
& 
\sum_{\phpol\ \ll1,\ll2} 
\int {d^{4-2\e}\ell\over (2\pi)^{4-2\e}}\;{i\over\ell^2}
  \Ctree_{-\sigma}((\ell+a+b)^{-\ll2},-\ell^{-\ll1}){i\over (\ell+k_a+k_b)^2}
\cr &\hskip 40mm\times
      A_{4}^\tree((-\ell-a-b)^\ll2,a^\la,b^\lb,\ell^\ll1)
}\eqn\OneLoopSplittingAmplitude$$
The restriction to physical polarizations is important; it will
give rise to transverse projection operators inside the loop.  The
defining equation is depicted graphically in fig.~\use\OneLoopSplittingFigure.

If we now combine all cuts, we find in agreement with the known
result~[\use\SusyFour], also shown in \fig\OneLoopFactorizationFigure,
$$\eqalign{
 A_n^{\oneloop}&(1,\ldots,a^\la,b^\lb,\ldots,n) 
\;{\buildrel a \parallel b\over{\relbar\mskip-1mu\joinrel\longrightarrow}}\cr
&\sum_{\phpol\ \sigma}  \biggl(
  \Split^\tree_{-\sigma}(a^{\la},b^{\lb})\,
      A_{n-1}^\oneloop(1,\ldots,(a+b)^\sigma,\ldots,n)
\cr &\hskip 20mm  
  +\Split^\oneloop_{-\sigma}(a^\la,b^\lb)\,
      A_{n-1}^\tree(1,\ldots,(a+b)^\sigma,\ldots,n) \biggr) \;.
}\eqn\OneLoopCollinear$$

This completes the proof of the universality of 
collinear factorization at one loop.  Unlike the proof
of Bern and Chalmers, it provides a compact formula
for computing the one-loop splitting amplitudes.  (Explicit
computations with this formula will be presented 
elsewhere~[\ref\KU{D. A. Kosower and P. Uwer, in preparation}].)
As we shall see in following sections, it also generalizes nicely
beyond one-loop amplitudes.

\LoadFigure\OneLoopFactorizationFigure
{\baselineskip 13 pt
\noindent\narrower\ninerm
A schematic depiction of the collinear factorization of one-loop
amplitudes.
}  {\epsfxsize 4.2 truein}{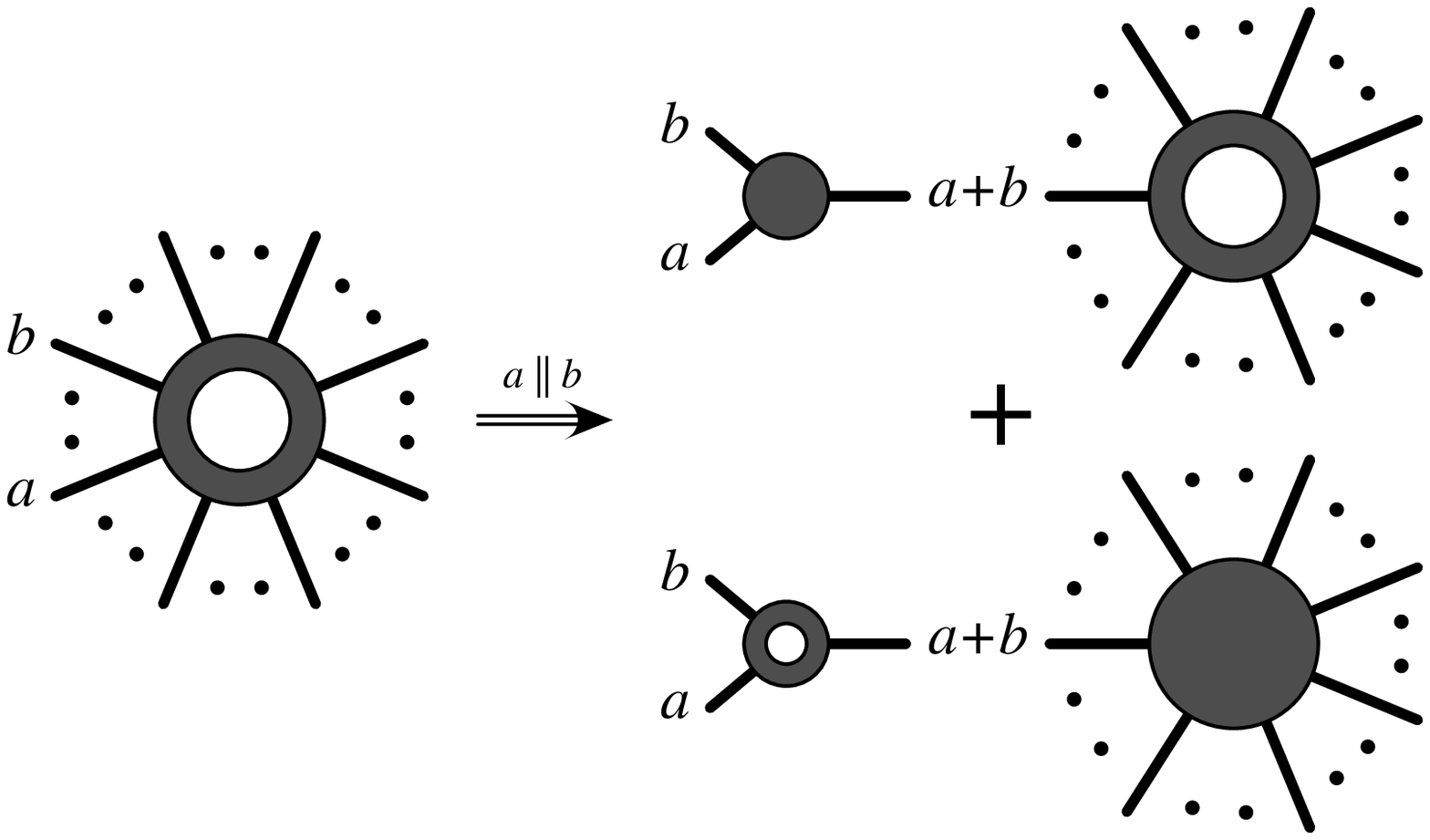}{}

\section{Two-Loop Collinear Factorization}
\vskip 10pt

The analysis in the previous sections extends readily to two-loop 
amplitudes.  At two loops or beyond, leading-color amplitudes (or equivalently,
 planar color-ordered amplitudes) do not suffice to construct the full
amplitude.
In the following, I will consider explicitly only leading-color amplitudes,
though the analysis can be extended to subleading-color ones.

The basic equation~(\use\BasicCut) of the unitarity-based method has
an analog at two- (or higher-) loop order.  The sum over cuts now
includes not only two-particle cuts but three-particle cuts
as well.  The amplitudes on either side of the cut are no longer
just tree amplitudes, but may be loop amplitudes as well,
$$\eqalign{
&\LP A_n^{\twoloop}(1,\ldots,a,b,\ldots,n) \RV_{t_{c\cdots d} {\rm\ cut}}
=\cr
&\hskip 5mm \int d\LIPS^{4-2\e}(\ell_1,\ell_2)\;
   A_{n-m+2}^\oneloop(\ell_1,c,\ldots,d,\ell_2) 
      A_{m+2}^\tree(-\ell_2,d+1,\ldots,a,b,\ldots,c-1,-\ell_1)\cr
&\hskip 10mm +\int d\LIPS^{4-2\e}(\ell_1,\ell_2)\;
   A_{n-m+2}^\tree(\ell_1,c,\ldots,d,\ell_2) 
      A_{m+2}^\oneloop(-\ell_2,d+1,\ldots,a,b,\ldots,c-1,-\ell_1)\cr
&\hskip 10mm +\int d\LIPS^{4-2\e}(\ell_1,\ell_2,\ell_3)\;
   A_{n-m+3}^\tree(\ell_1,c,\ldots,d,\ell_2,\ell_3) \cr
&\hskip 20mm \times
      A_{m+3}^\tree(-\ell_3,-\ell_2,d+1,\ldots,a,b,\ldots,c-1,-\ell_1)\,.\cr
}\eqn\twoLoopTypeNS$$
Note that in the three-particle
cuts, all the $\ell_i$ must be adjacent in order to obtain a leading-color
(planar) contribution; in the labelling used here,
 $\ell_3$ is actually the middle leg crossing the cut.  The sum over
intermediate particle types and physical polarizations should again
be understood implicitly.

The arguments on the ordering of limits given in section~\use\LimitOrderSection\ 
are in fact quite general, and apply to higher-loop amplitudes as well.
We may thus take the singular limit first, in all channels other than that
of the singular invariant $s_{ab}$.
To do so, we must use both the tree-level collinear limit~(\use\CollinearA)
as well as the one-loop limit~(\use\OneLoopCollinear), whereupon we
find that the above expression has the following limit,
$$\eqalign{
&\sum_{\phpol\ \sigma} \Ctree_{-\sigma}(a^\la,b^\lb)\,\LB
  \int d\LIPS^{4-2\e}(\ell_1,\ell_2)\;
   A_{n-m+2}^\oneloop(\ell_1,c,\ldots,d,\ell_2) \RP
\cr &\hskip 35mm\times
      A_{m+1}^\tree(-\ell_2,d+1,\ldots,(a+b)^\sigma,\ldots,c-1,-\ell_1)\cr
&\hskip 25mm + \int d\LIPS^{4-2\e}(\ell_1,\ell_2)\;
   A_{n-m+2}^\tree(\ell_1,c,\ldots,d,\ell_2) 
\cr &\hskip 35mm\times
      A_{m+1}^\oneloop(-\ell_2,d+1,\ldots,(a+b)^\sigma,\ldots,c-1,-\ell_1)\cr
&\hskip 25mm + \int d\LIPS^{4-2\e}(\ell_1,\ell_2,\ell_3)\;
   A_{n-m+3}^\tree(\ell_1,c,\ldots,d,\ell_2,\ell_3) 
\cr &\hskip 35mm\times\LP\vphantom{\int }
    A_{m+2}^\tree(-\ell_3,-\ell_2,d+1,\ldots,(a+b)^\sigma,\ldots,c-1,-\ell_1)
    \RB\cr
&\hskip 5mm +\sum_{\phpol\ \sigma} \Cone_{-\sigma}(a^\la,b^\lb)\,
    \int d\LIPS^{4-2\e}(\ell_1,\ell_2)\;
   A_{n-m+2}^\tree(\ell_1,c,\ldots,d,\ell_2) 
\cr &\hskip 45mm\times
      A_{m+1}^\tree(-\ell_2,d+1,\ldots,(a+b)^\sigma,\ldots,c-1,-\ell_1)\cr
&=
\sum_{\phpol\ \sigma} \Ctree_{-\sigma}(a^\la,b^\lb)\,
  \LP A_{n-1}^\twoloop(1,\ldots,(a+b)^\sigma,\ldots,n) 
  \RV_{t_{c\cdots d} {\rm\ cut}}\cr
&\hskip 13mm +\sum_{\phpol\ \sigma} \Cone_{-\sigma}(a^\la,b^\lb)\,
  \LP A_{n-1}^\oneloop(1,\ldots,(a+b)^\sigma,\ldots,n) 
  \RV_{t_{c\cdots d} {\rm\ cut}}\cr
}\eqn\TLanyChannel$$

We must again consider the singular channel specially.  The cut in
that channel is,
$$\eqalign{
&\LP A_n^{\twoloop}(1,\ldots,a,b,\ldots,n) \RV_{s_{ab} {\rm\ cut}}
=\cr
&\hskip 5mm \int d\LIPS^{4-2\e}(\ell_1,\ell_2)\;
   A_{n}^\oneloop(-\ell_1,b+1,\ldots,a-1,-\ell_2) 
      A_{4}^\tree(\ell_2,a,b,\ell_1)\cr
&\hskip 10mm +\int d\LIPS^{4-2\e}(\ell_1,\ell_2)\;
   A_{n}^\tree(-\ell_1,b+1,\ldots,a-1,-\ell_2) 
      A_{4}^\oneloop(\ell_2,a,b,\ell_1)\cr
&\hskip 10mm +\int d\LIPS^{4-2\e}(\ell_1,\ell_2,\ell_3)\;
   A_{n+1}^\tree(-\ell_1,b+1,\ldots,a-1,-\ell_2,-\ell_3) 
      A_{5}^\tree(\ell_3,\ell_2,a,b,\ell_1)\cr
}\eqn\TLsingularChannel$$

Once again, momentum conservation forces all legs crossing the cut to
become collinear.  For the two particle cuts, 
this is just the statement that $s_{\ell_1\ell_2}\rightarrow 0$, with
the $\ell_i$ massless.  For
the three-particle cuts, the three-particle invariant vanishes,
$t_{\ell_1\ell_2\ell_3}\rightarrow 0$.  Because the $\ell_i$ are
physical (in a $D$-dimensional sense), they all have positive energies;
the vanishing of the three-particle invariant thus forces the separate
two-particle invariants to vanish as well, $s_{\ell_i\ell_j}\rightarrow 0$.

The behavior of tree amplitudes when three color-adjacent external
particles become collinear
is governed by analogs of the collinear splitting amplitude,
$$\eqalign{
&A_{n+2}^\tree(\ldots,a^\la,b^\lb,c^\lc\ldots) 
   \inlimit^{k_a \tcdot k_b, k_a\tcdot k_c, k_b\tcdot k_c\rightarrow 0}
\cr &\hskip 15mm
 \sum_{\phpol\ \sigma}
 \Ctree_{-\sigma}(a^\la,b^\lb,c^\lc) 
    A_{n}^\tree(\ldots,(a+b+c)^\sigma,\ldots)\,,\cr
}\eqn\DoubleCollinear$$
valid when $n\geq 4$.  The squares
of such {\it double-collinear\/} splitting amplitudes have been computed
by Glover and Campbell~[\use\EnglishDoubleCollinear] and by Catani and 
Grazzini~[\use\ItalianDoubleCollinear].

As in the one-loop case, we cannot factorize the four- or five-point
amplitudes on the right-hand side of the cut.  We can, however, factorize
the amplitudes on the left-hand side, obtaining
$$\eqalign{
& \sum_{{\phpol\atop\sigma, \ll1,\ll2}}
   A_{n-1}^\oneloop((a+b)^\sigma,b+1,\ldots,a-1) 
\cr &\hskip 30mm\times
   \int d\LIPS^{4-2\e}(\ell_1,\ell_2)\;
    \Ctree_{-\sigma}(-\ell_2^\ll2,-\ell_1^\ll1)
      A_{4}^\tree(\ell_2,a,b,\ell_1)\cr
&\hskip 4mm +\sum_{{\phpol\atop \sigma, \ll1,\ll2}}
   A_{n-1}^\tree((a+b)^\sigma,b+1,\ldots,a-1) 
\cr &\hskip 30mm\times
   \int d\LIPS^{4-2\e}(\ell_1,\ell_2)\;
    \Cone_{-\sigma}(-\ell_2^\ll2,-\ell_1^\ll1)
      A_{4}^\tree(\ell_2,a,b,\ell_1)\cr
&\hskip 4mm +\sum_{{\phpol\atop \sigma, \ll1,\ll2}}
   A_{n-1}^\tree((a+b)^\sigma,b+1,\ldots,a-1) 
\cr &\hskip 30mm\times
     \int d\LIPS^{4-2\e}(\ell_1,\ell_2)\;
    \Ctree_{-\sigma}(-\ell_2^\ll2,-\ell_1^\ll1)
      A_{4}^\oneloop(\ell_2,a,b,\ell_1)\cr
&\hskip 4mm +\sum_{{\phpol\atop \sigma, \ll1,\ll2, \ll3}}
   \hskip -3mm A_{n-1}^\tree((a+b)^\sigma,b+1,\ldots,a-1) 
\cr &\hskip 30mm\times
  \int d\LIPS^{4-2\e}(\ell_1,\ell_2,\ell_3)\;
    \Ctree_{-\sigma}(-\ell_2^\ll2,-\ell_3^\ll3,-\ell_1^\ll1)
      \rlap{$A_{5}^\tree(\ell_3,\ell_2,a,b,\ell_1)\,.$}\cr
}\anoneqn$$

We recognize
$$
   \int d\LIPS^{4-2\e}(\ell_1,\ell_2)\;
    \Ctree_{-\sigma}(-\ell_2^\ll2,-\ell_1^\ll1)
      A_{4}^\tree(\ell_2,a,b,\ell_1)
\anoneqn$$
as the cut of the one-loop splitting function, so that eqn.~(\use\TLsingularChannel)
can be written as
$$\eqalign{
&\sum_{{\phpol\atop \sigma, \ll1,\ll2}}
   \LP \Cone_{-\sigma}(a^\la,b^\lb)\RV_{s_{ab} {\rm\ cut}}
   A_{n-1}^\oneloop((a+b)^\sigma,b+1,\ldots,a-1) \cr
&\hskip 5mm +\sum_{{\phpol\atop \sigma, \ll1,\ll2}}
   \LP \Ctwo_{-\sigma}(a^\la,b^\lb)\RV_{s_{ab} {\rm\ cut}}
   A_{n-1}^\tree((a+b)^\sigma,b+1,\ldots,a-1) \,,
}\eqn\TLsingularChannelFinal$$
where I have introduced a new quantity, the two-loop splitting
amplitude $\Ctwo$.  It is defined by its cuts,
$$\eqalign{
   \Ctwo_{-\sigma}&\LP(a^\la,b^\lb)\RV_{s_{ab} {\rm\ cut}}\cr
&= \sum_{{\phpol\atop \sigma, \ll1,\ll2}}
   \int d\LIPS^{4-2\e}(\ell_1,\ell_2)\;\LB
    \Cone_{-\sigma}(-\ell_2^\ll2,-\ell_1^\ll1)
      A_{4}^\tree(\ell_2,a,b,\ell_1)\RP\cr
&\hskip 50mm\LP + \Ctree_{-\sigma}(-\ell_2^\ll2,-\ell_1^\ll1)
      A_{4}^\oneloop(\ell_2,a,b,\ell_1)\RB\cr
&\hskip 5mm +\sum_{{\phpol\atop \sigma, \ll1,\ll2, \ll3}}
  \int d\LIPS^{4-2\e}(\ell_1,\ell_2,\ell_3)\;
    \Ctree_{-\sigma}(-\ell_2^\ll2,-\ell_3^\ll3,-\ell_1^\ll1)
      A_{5}^\tree(\ell_3,\ell_2,a,b,\ell_1)\,.\cr
}\eqn\Ctwodef$$
The splitting amplitude has no cuts in other channels.
Because the cut in this
channel is taken to all orders in $\e$, this equation suffices to
reconstruct it completely.

\LoadFigure\TwoLoopFactorizationFigure
{\baselineskip 13 pt
\noindent\narrower\ninerm
A schematic depiction of the collinear factorization of two-loop
amplitudes.
}  {\epsfxsize 4.2 truein}{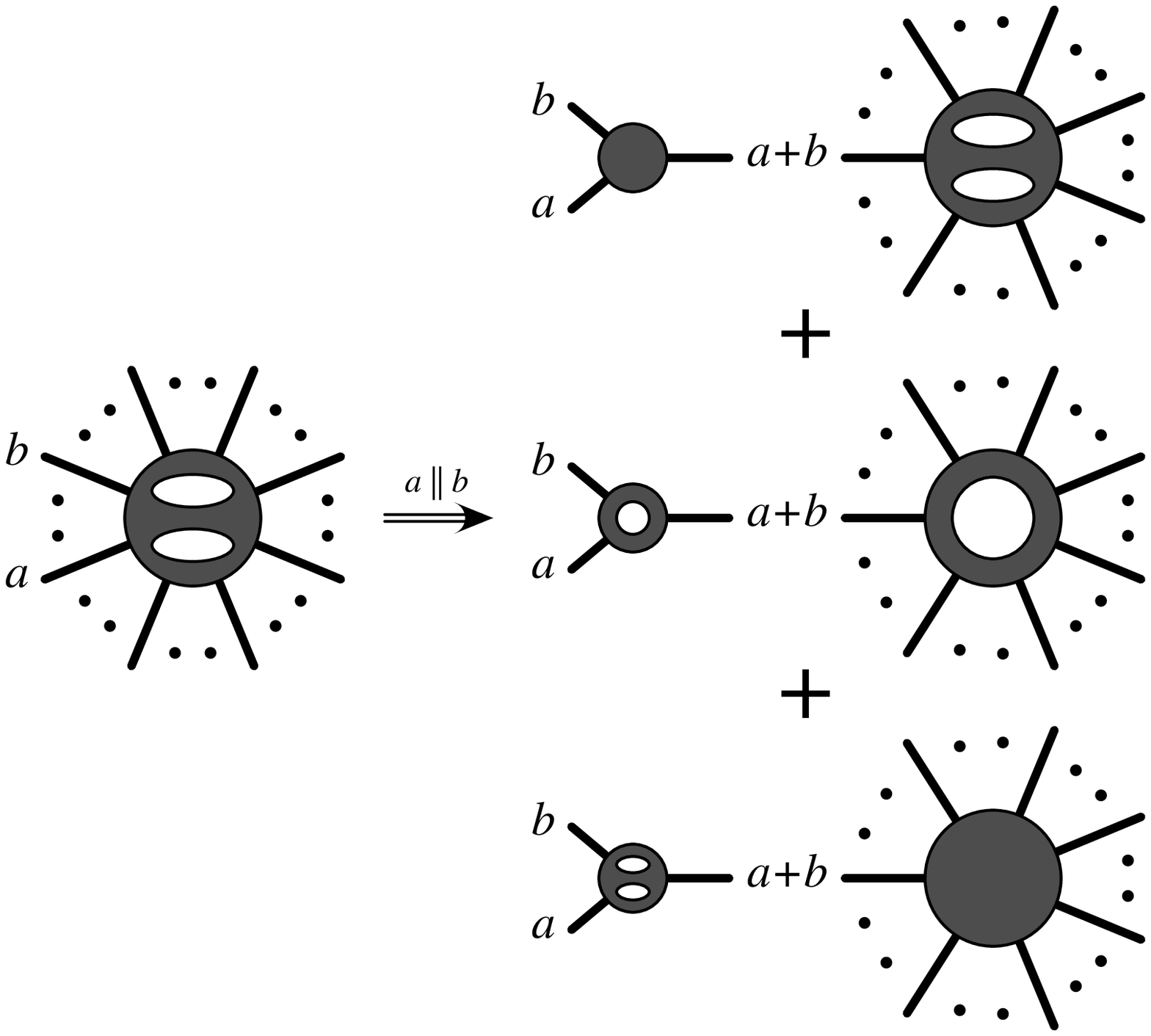}{}

The first term in eqn.~(\use\TLsingularChannelFinal) repoduces
the second term on the last line of eqn.~(\use\TLanyChannel); combining all channels,
we finally obtain the collinear behavior of a leading-color two-loop
amplitude, also shown schematically in fig.~\use\TwoLoopFactorizationFigure,
$$\eqalign{ A_n^\twoloop&(1,\ldots,a^\la,b^\lb,\ldots,n)
   \inlimit^{k_a \cdot k_b\rightarrow 0}\cr
&\hskip 5mm\sum_{\phpol\ \sigma} \LB \Ctree_{-\sigma}(a^\la,b^\lb)\,
   A_{n-1}^\twoloop(1,\ldots,(a+b)^\sigma,\ldots,n) \RP\cr
&\hskip 19mm + \Cone_{-\sigma}(a^\la,b^\lb)\,
   A_{n-1}^\oneloop(1,\ldots,(a+b)^\sigma,\ldots,n) 
  \vphantom{\sum_{\phpol\ \sigma}} \cr
&\hskip 19mm\LP + \Ctwo_{-\sigma}(a^\la,b^\lb)\,
  A_{n-1}^\tree(1,\ldots,(a+b)^\sigma,\ldots,n) \RB\,.
  \vphantom{\sum_{\sigma=\pm}} \cr
}\anoneqn$$

\section{One-Loop Corrections to the Double-Collinear Splitting Amplitude}
\vskip 10pt

As three external legs become simultaneously collinear in a one-loop
amplitude, we expect a factorization analogous to that at tree level,
eqn.~(\use\DoubleCollinear).  We can derive it using an approach similar
to that in previous sections.  Much of the discussion will carry over,
but there are certain differences worthy of closer examination.

The major difference is that the double-collinear splitting amplitude
is no longer a function of a sole invariant; rather, it depends on
the two neighboring two-particle invariants $s_{ab}$ and $s_{bc}$, as well as the
overall three-particle invariant $t_{abc}$.  All of these invariants vanish
in the limit.  (I take the order of the external legs to be $a,b,c$.)
As
discussed by Campbell and Glover~[\use\EnglishDoubleCollinear],
the tree-level double-collinear splitting amplitude contains
all terms that are singular in two of the invariants.  (Terms singular
in only one invariant would eventually give rise to 
a vanishing contribution to physical processes.)
At one loop, we will acquire additional epsilonic powers of the
invariants (and possibly more complicated analytic functions of
their ratios), but the basic requirement of having integral
or half-integral power singularities in two of the invariants remains.

As in the simple splitting amplitudes, these integral powers can arise
either from amplitudes on either side of the cut, or else from the
loop integral whose cut we are taking.  In the former case, all three
of the collinear legs must be on the same side of the cut.  In the
latter case, an inverse power in a given invariant also implies the
integral function must contain a cut in that invariant.  Let us focus
on the inverse powers of $t_{abc}$.
Two of the
possible pole combinations --- in $(s_{ab},t_{abc})$ and in $(s_{bc},t_{abc})$
--- manifestly require that the loop integral have a cut in $t_{abc}$.
The third combination $(s_{ab},s_{bc})$ also requires such a cut.  To 
see this, note that it must have cuts in both $s_{ab}$ and $s_{bc}$,
and thus must have uncancelled propagators between legs $a-1$ and $a$,
and likewise $c$ and $c+1$.  These two uncancelled propagators force
the integral to have a cut in the three-particle channel as well.

The set of cuts we must consider is thus exactly analogous to that
for the simple collinear splitting amplitude: all legs must
be on one side of the cut.  Contributions missed by ignoring cuts
that separate the collinear legs will be be picked up in the
primary singular channel, where the three legs will sit alone on one side
of the cut.  The remaining arguments about the order of limits
go through as before: in channels other than the singular channel,
we can take the singular limit before performing the cut, in the
singular channel we must tread carefully.  (These arguments also carry
over to general multi-collinear splitting amplitudes.)

In a non-singular channel, the limit of eqn.~(\use\typeNS) again
produces a tree-level splitting amplitude,
$$\eqalign{
&\LP A_n^{\oneloop}(1,\ldots,a,b,c,\ldots,n) \RV_{t_{d\cdots f} {\rm\ cut}}
   \inlimit^{k_a \tcdot k_b, k_a\tcdot k_c, k_b\tcdot k_c\rightarrow 0}
\cr &\hskip 10mm 
\sum_{\phpol\ \sigma} \Ctree_{-\sigma}(a^\la,b^\lb,c^\lc)\,
   \int d\LIPS^{4-2\e}(\ell_1,-\ell_2)\;
   \cr
&\hskip 15mm\times
   A_{n-m+2}^\tree(\ell_1,d,\ldots,f,-\ell_2) 
      A_{m}^\tree(\ell_2,f\!+\!1,\ldots,(a+b+c)^\sigma,\ldots,d\!-\!1,-\ell_1)\cr
&= \sum_{\phpol\ \sigma} \Ctree_{-\sigma}(a^\la,b^\lb,c^\lc)\,
\LP A_{n-2}^{\oneloop}(1,\ldots,(a+b+c)^\sigma,\ldots,n) 
           \RV_{t_{d\cdots f} {\rm\ cut}} \,.
\cr}\eqn\DoubleCollinearB$$

In the three-particle singular channel, the cut is
$$\eqalign{
&\LP A_n^{\oneloop}(1,\ldots,a,b,c,\ldots,n) \RV_{t_{abc} {\rm\ cut}}
=\cr
&\hskip 10mm \sum_{\phpol \lambda_{1,2}} \int d\LIPS^{4-2\e}(\ell_1,\ell_2)\;
   A_{n-1}^\tree(-\ell_1^{-\ll1},c\!+\!1,\ldots,a\!-\!1,-\ell_2^{-\ll2}) 
      A_{5}^\tree(\ell_2^\ll2,a,b,c,\ell_1^\ll1)\,.\cr
}\eqn\DCtypeS$$
The vanishing of the three-particle invariant forces the vanishing of
the invariant of the legs crossing the cut, $s_{\ell_1\ell_2}$, and
we can again use this to factorize the amplitude on the left-hand side,
so the cut becomes
$$\eqalign{
&\sum_{{\phpol\atop\sigma,\ll1,\ll2}} 
     A_{n-2}^\tree((a+b+c)^\sigma,c\!+\!1,\ldots,a\!-\!1)\,
\cr&\hskip 20mm\times
\int d\LIPS^{4-2\e}(\ell_1,\ell_2)\;
   \Ctree_{-\sigma}(-\ell_2^{-\ll2},-\ell_1^{-\ll1})\,
      A_{5}^\tree(\ell_2^\ll2,a^\la,b^\lb,c^\lc,\ell_1^\ll1)\cr
&\equiv \sum_{\phpol\ \sigma} A_{n-2}^\tree((a+b+c)^\sigma,c\!+\!1,\ldots,a\!-\!1)\,
\LP\Cone_{-\sigma}(a^\la,b^\lb,c^\lb)\RV_{t_{abc} \rm\, cut}\,,\cr
}\eqn\DCtypeSb$$
where the latter equation defines the one-loop double splitting amplitude.

Although this splitting amplitude will also have cuts in the two-particle
invariants within the collinear set, as discussed above it must have a cut
in the three-particle channel, and hence this cut suffices to reconstruct it fully,
$$\eqalign{
\Cone_{-\sigma}&(a^\la,b^\lb,c^\lc) = \cr
& 
\sum_{\phpol\ \ll1,\ll2} 
\int {d^{4-2\e}\ell\over (2\pi)^{4-2\e}}\;{i\over\ell^2}
  \Ctree_{-\sigma}((\ell+a+b+c)^{-\ll2},-\ell^{-\ll1}){i\over (\ell+k_a+k_b+k_c)^2}
\cr &\hskip 40mm\times
      A_{5}^\tree((-\ell-a-b)^\ll2,a^\la,b^\lb,c^\lc,\ell^\ll1)
}\eqn\OneLoopDoubleCollinearSplittingAmplitude$$
The restriction to physical polarizations will once again
give rise to transverse projection operators inside the loop.

Combining all cuts, we find a double-collinear factorization formula very
similar to the simple collinear one,
$$\eqalign{
 A_n^{\oneloop}&(1,\ldots,a^\la,b^\lb,c^\lc\ldots,n) 
\;{\buildrel a \parallel b\parallel c\over{\relbar\mskip-1mu\joinrel\longrightarrow}}\cr
&\sum_{\phpol\ \sigma}  \biggl(
  \Split^\tree_{-\sigma}(a^{\la},b^{\lb},c^\lc)\,
      A_{n-2}^\oneloop(1,\ldots,(a+b+c)^\sigma,\ldots,n)
\cr &\hskip 20mm  
  +\Split^\oneloop_{-\sigma}(a^\la,b^\lb,c^\lc)\,
      A_{n-2}^\tree(1,\ldots,(a+b+c)^\sigma,\ldots,n) \biggr) \;.
}\eqn\OneLoopDoubleCollinear$$

\section{Collinear Factorization to All Orders}
\vskip 10pt

In this section, I give a proof of collinear factorization to
all orders, generalizing further the results of the previous sections.
As in the two-loop case, I will restrict attention to leading-color
amplitudes, that is to coefficients of $N^l$ times the tree-level color structure
in an $l$-loop amplitude.

The general cut formula here is
$$\eqalign{
&\LP A_n^{l\lloop}
(1,\ldots,a_1,a_2,\ldots,a_c,\ldots,n) \RV_{t_{d\cdots f} {\rm\ cut}}
=\cr
&\hskip 5mm
\sum_{j=2}^{l+1} \int d\LIPS^{4-2\e}(\ell_1,\ldots,\ell_j)\;\cr
&\hskip 10mm\times
   \sum_{k=0}^{l+1-j}
   A_{n-m+j}^{k\lloop}(d,\ldots,f,\ell_1,\ldots,\ell_j) 
   \,A_{m+j}^{(l+1-j-k)\lloop}
          (f\!+\!1,\ldots,a,b,\ldots,d\!-\!1,-\ell_j,\ldots,-\ell_1)\cr
}\eqn\allLoopTypeNS$$
so long as the $a_i$ all end up on one side of the cut.  (As in earlier sections,
we can ignore contributions where different $a_i$ end up on opposite sides of the cut.)
The outer sum is on $j$-particle cuts, the inner one on the different $j$-particle
cuts; $m$ is as in previous sections, the number of external momenta on the far
side of the cut; and `0-loop' means `tree'.  
 The sum over
intermediate particle types and physical polarizations is, as always, implicit.

The proof will proceed by induction, in fact a double induction, over both
the number of loops and the number of color-adjacent legs becoming collinear
simultaneously.  Denote by $S_c$ the set of invariants built out of consecutive
momenta in $\{a_1,\ldots,a_c\}$, and indicate the multiple collinear
limit by $S_c\rightarrow 0$ (that is, all of the invariants are supposed to vanish).

Now assume that for $r<l$ and $c<n-2$, that we have the following factorization,
$$\eqalign{ A_n^{r\lloop}&(1,\ldots,a_1^\ll1,a_2^\ll2,\ldots,a_c^\ll{c},\ldots,n)
   \inlimit^{S_c\rightarrow 0}\cr
&\sum_{\phpol\ \sigma} \sum_{v=0}^r
 \Split^{v\lloop}_{-\sigma}(a_1^\ll1,a_2^\ll2,\ldots,a_c^\ll{c})\,
   A_{n-c+1}^{(r-v)\lloop}(1,\ldots,(a_1+\ldots+a_c)^\sigma,\ldots,n) \,,
}\eqn\InductionAnsatz$$
The starting point for the induction --- $r=1$ and $c=2$ --- is the one-loop
factorization proven in section~\OneLoopSingularSection.  In addition, I will make
use of the tree-level multi-collinear analog to eqn.~(\use\DoubleCollinear),
which can again be derived either from the Berends--Giele recurrence relation 
or from the string representation.

In all cuts except that in the primary singular channel 
($K_c^2 = (k_{a_1}+\ldots+k_{a_c})^2$),
we then find using the above assumption, that
$$\eqalign{
&\LP A_n^{l\lloop}
(1,\ldots,a_1^\ll1,a_2^\ll2,\ldots,a_c^\ll{c},\ldots,n) \RV_{t_{d\cdots f} {\rm\ cut}}
   \inlimit^{S_c\rightarrow 0}\cr
&\hskip 5mm
\sum_{j=2}^{l+1} \int d\LIPS^{4-2\e}(\ell_1,\ldots,\ell_j)\;
   \sum_{k=0}^{l+1-j}
   A_{n-m+j}^{k\lloop}(d,\ldots,f,\ell_1,\ldots,\ell_j) 
\cr&\hskip 10mm\times
\sum_{\phpol\ \sigma} \sum_{v=0}^{l+1-j-k}
 \Split^{v\lloop}_{-\sigma}(a_1^\ll1,a_2^\ll2,\ldots,a_c^\ll{c})
\cr&\hskip 25mm\times
   \,A_{m+j-c+1}^{(l+1-j-k-v)\lloop}
          (f\!+\!1,\ldots,(a_1+\ldots+a_c)^\sigma,\ldots,
              d\!-\!1,-\ell_j,\ldots,-\ell_1)\,.\cr
}\eqn\allOrdersTypeNS$$
Interchanging the summations,
$$\eqalign{
\sum_{j=2}^{l+1} \sum_{k=0}^{l+1-j} \sum_{v=0}^{l+1-j-k}
 = \sum_{j=2}^{l+1} \sum_{v=0}^{l+1-j} \sum_{k=0}^{l+1-j-v}
=\sum_{v=0}^{l-1} \sum_{j=2}^{l+1-v} \sum_{k=0}^{l+1-j-v}\,,
}\anoneqn$$
we can transform eqn.~(\use\allOrdersTypeNS) to obtain
$$\eqalign{
&\LP A_n^{l\lloop}
(1,\ldots,a_1^\ll1,a_2^\ll2,\ldots,a_c^\ll{c},\ldots,n) \RV_{t_{d\cdots f} {\rm\ cut}}
   \inlimit^{S_c\rightarrow 0}\cr
&\hskip 5mm
\sum_{\phpol\ \sigma}\,
\sum_{v=0}^{l-1} 
 \Split^{v\lloop}_{-\sigma}(a_1^\ll1,a_2^\ll2,\ldots,a_c^\ll{c})
\sum_{j=2}^{l-v+1} \int d\LIPS^{4-2\e}(\ell_1,\ldots,\ell_j)\;
\cr&\hskip 10mm\times
   \sum_{k=0}^{l-v+1-j}
   A_{n-m+j}^{k\lloop}(d,\ldots,f,\ell_1,\ldots,\ell_j) 
\cr&\hskip 20mm\times
   \,A_{m+j-c+1}^{(l+1-j-k-v)\lloop}
          (f\!+\!1,\ldots,(a_1+\ldots+a_c)^\sigma,\ldots,
           d\!-\!1,-\ell_j,\ldots,-\ell_1)\,\cr
&\hskip 5mm
=\sum_{\phpol\ \sigma}
\sum_{v=0}^{l-1} 
 \Split^{v\lloop}_{-\sigma}(a_1^\ll1,a_2^\ll2,\ldots,a_c^\ll{c})
\,\LP A_{n-c+1}^{(l-v)\lloop}
(1,\ldots,(a_1+\ldots+a_c)^\sigma,\ldots,n) \RV_{t_{d\cdots f} {\rm\ cut}}
}\anoneqn$$
which is exactly eqn.~(\use\InductionAnsatz) in this channel, but now
for $r=l$.  It again holds for all $c<n-2$.

In the primary singular channel, we obtain in the limit,
$$\eqalign{
&\sum_{j=2}^{l+1} \int d\LIPS^{4-2\e}(\ell_1,\ldots,\ell_j)\;
   \sum_{k=0}^{l+1-j}
\,\sum_{\phpol\ \sigma}\sum_{v=0}^{k} 
 \Split^{v\lloop}_{-\sigma}(\ell_1,\ldots,\ell_j)
\cr&\hskip 10mm\times
   A_{n-c+1}^{(k-v)\lloop}(a_c\!+\!1,\ldots,a_1\!-\!1,(a_1+\ldots+a_c)^\sigma) 
\cr&\hskip 15mm\times
   \,A_{c+j}^{(l+1-j-k)\lloop}
          (a_1^\ll1,a_2^\ll2,\ldots,a_c^\ll{c},-\ell_j,\ldots,-\ell_1)\,\cr
&=\sum_{k=0}^{l-1} 
\,\sum_{\phpol\ \sigma}\,\sum_{v=0}^{k} 
   A_{n-c+1}^{(k-v)\lloop}(a_c\!+\!1,\ldots,a_1\!-\!1,(a-1+\ldots+a_c)^\sigma) \cr
&\hskip 10mm\times
\sum_{j=2}^{l+1-k} \int d\LIPS^{4-2\e}(\ell_1,\ldots,\ell_j)\;
 \Split^{v\lloop}_{-\sigma}(\ell_1,\ldots,\ell_j)
\cr &\hskip 20mm\times
   \,A_{c+j}^{(l+1-j-k)\lloop}
          (a_1^\ll1,a_2^\ll2,\ldots,a_c^\ll{c},-\ell_j,\ldots,-\ell_1)\,\cr
}\eqn\allOrdersTypeS$$
which gives us an explicit formula allowing the reconstruction of
the $l$-loop multiparticle splitting amplitude
in eqn.~(\use\InductionAnsatz) as the coefficient of $A^\tree_{n-c+1}$,
$$\eqalign{
&\LP \Split^{l\lloop}_{-\sigma}(a_1^\ll1,a_2^\ll2,\ldots,a_c^\ll{c})
   \RV_{K_c^2 {\rm\ cut}}
\cr &\hskip 10mm=\sum_{k=0}^{l-1} 
\,\sum_{j=2}^{l+1-k} 
\sum_{\phpol\ \sigma_i}\int d\LIPS^{4-2\e}(\ell_1,\ldots,\ell_j)\;
%\cr&\hskip 25mm\times
 \Split^{k\lloop}_{-\sigma}(\ell_1^{-\sigma_1},\ldots,\ell_j^{-\sigma_j})
\cr&\hskip 30mm\times
   \,A_{c+j}^{(l+1-j-k)\lloop}
  (a_1^\ll1,a_2^\ll2,\ldots,a_c^\ll{c},-\ell_j^{\sigma_j},\ldots,-\ell_1^{\sigma_1})\,\cr
}\anoneqn$$

Combining the two kinds of channels, we obtain eqn.~(\use\InductionAnsatz)
uniformly for $r=l$, thereby proving the desired result.

\section{Conclusions}
\vskip 10pt

The square-root factorization of amplitudes in the collinear limit is a
striking feature of gauge theories, yet one that hides subtleties.  The
subtleties, and the complexity of a conventional diagrammatic approach
to the problem, are associated with the presence of infrared singularities.
The naive intuition of computing only those diagrams associated, for example,
 with fig.~\use\OneLoopFactorizationFigure, is not quite right.  This is
demonstrated graphically by the presence of dilogarithms in the quark-gluon
splitting amplitudes~[\use\SusyFour,\use\qqggg], which cannot arise from
massless three-point integrals at one loop.  The unitarity-based proof
given in this paper circumvents these complexities, and gives as well explicit
formul\ae\ for computing the splitting amplitudes.

\vskip 20pt

I thank P. Uwer, Z. Bern, and L. Dixon for helpful comments.

\listrefs
\bye